\documentclass[]{JHEP3}

\title{\bf Counterterms in Dimensionally Continued AdS Gravity}

\author{Olivera Mi\v{s}kovi\'c $^{a,\,b}$ and Rodrigo Olea $^c$\medskip \\
$^a$Instituto de F\'{\i}sica, P. Universidad
Cat\'{o}lica de Valpara\'{\i}so, Casilla 4059,\\
Valpara\'{\i}so, Chile.\\
$^b$Centro Multidisciplinar de Astrof\'\i sica (CENTRA),
Departamento de F\'\i sica,\\
Instituto Superior T\'ecnico, Universidade T\'ecnica de Lisboa,\\
Av.Rovisco Pais 1, 1049-001 Lisboa, Portugal.\\
$^{c}$INFN, Sezione di Milano, Via Celoria 16, I-20133,
Milano, Italy.\smallskip \\
E-mail: \email{olivera.miskovic@ucv.cl},
\email{rodrigo.olea@mi.infn.it,}}

\preprint{IFUM-896-FT\\ \arXivid{0706.4460}}

\abstract{We revise two regularization mechanisms for Lovelock
gravity with AdS asymptotics. The first one corresponds to the
Dirichlet counterterm method, where local functionals of the
boundary metric are added to the bulk action on top of a
Gibbons-Hawking-Myers term that defines the Dirichlet problem in
gravity. The generalized Gibbons-Hawking term can be found in any
Lovelock theory following the Myers' procedure to achieve a
well-posed action principle for a Dirichlet boundary condition on
the metric, which is proved to be equivalent to the Hamiltonian
formulation for a radial foliation of spacetime. In turn, a closed
expression for the Dirichlet counterterms does not exist for a
generic Lovelock gravity. The second method supplements the bulk
action with boundary terms which depend on the extrinsic curvature
(Kounterterms), and whose explicit form is independent of the
particular theory considered.

In this paper, we use Dimensionally Continued AdS Gravity
(Chern-Simons-AdS in odd and Born-Infeld-AdS in even dimensions)
as a toy model to perform the first explicit comparison between
both regularization prescriptions. This can be done thanks to the
fact that, in this theory, the Dirichlet counterterms can be
readily integrated out from the divergent part of the Dirichlet
variation of the action.

The agreement between both procedures at the level of the boundary
terms suggests the existence of a general property of any
Lovelock-AdS gravity: intrinsic counterterms are generated as the
difference between the Kounterterm series and the corresponding
Gibbons-Hawking-Myers term.}


\begin{document}

\section{Introduction}

Lovelock gravity \cite{Lanczos-Lovelock} has recently attracted great
interest in theoretical physics as higher-curvature terms have been shown to
appear in the low-energy limit of String Theory as corrections to
Einstein-Hilbert action.

Lovelock gravity in $D=d+1$ dimensions is described by the action
\begin{equation}
I=\kappa \sum\limits_{p=0}^{[(D-1)/2]}\alpha _{p}\,I^{(p)}\,,
\label{Lovelock}
\end{equation}%
where $I^{(p)}$ corresponds to the dimensional continuations of $p$%
-dimensional Euler density, i.e.,
\begin{equation}
I^{(p)}=\int\limits_{M_{D}}\varepsilon _{A_{1}\cdots A_{D}}\,\hat{R}%
^{A_{1}A_{2}}\cdots \hat{R}^{A_{2p-1}A_{2p}}e^{A_{2p+1}}\cdots e^{A_{D}}\,,
\end{equation}%
that carries an arbitrary weight factor $\alpha _{p}${\bf \ }and $\kappa $
is a gravitational constant.{\bf \ }The vielbein $e^{A}=e_{\mu
}^{A}\,dx^{\mu }$ is related to the spacetime metric by $G_{\mu \nu }=\eta
_{AB}\,e_{\mu }^{A}\,e_{\nu }^{B}$, and $\hat{R}^{AB}=d\omega ^{AB}+\omega
^{AC}\omega _{C}^{\;\;B}$ is the Lorentz curvature associated to the spin
connection 1-form $\omega ^{AB}=\omega _{\mu }^{AB}\,dx^{\mu }$. The
curvature 2-form can be expressed in terms of the spacetime Riemann tensor
as $\hat{R}^{AB}=\frac{1}{2}\,\hat{R}_{\mu \nu }^{\alpha \beta }\,e_{\alpha
}^{A}\,e_{\beta }^{B}\,dx^{\mu }dx^{\nu }$. The sets $\{A,B,\ldots \}$ and $%
\{\mu ,\nu ,\ldots \}$ label tangent space and spacetime indices,
respectively. The tensorial equivalence of the action $I^{(p)}$ reads
\begin{equation}
I^{(p)}=-\frac{\left( D-2p\right) !}{2^{p}}\,\int\limits_{M_{D}}d^{D}x\,%
\sqrt{-G}\,\delta _{\left[ \mu _{1}\cdots \mu _{2p}\right] }^{\left[ \nu
_{1}\cdots \nu _{2p}\right] }\,\hat{R}_{\nu _{1}\nu _{2}}^{\mu _{1}\mu
_{2}}\cdots \hat{R}_{\nu _{2p-1}\nu _{2p}}^{\mu _{2p-1}\mu _{2p}}\,,
\label{p-Lovelock}
\end{equation}%
where the totally-antisymmetric Kronecker delta and its properties
are given in Appendix A. Because the action $I$ is a linear
combination of all dimensionally continued lower-dimensional Euler
densities, the derived equations of motion are at most of second
order in the metric, what frees
this theory from ghosts when expanded around a flat background \cite%
{ghost-free}. General covariance, together with second-order field
equations, are the basic features of General Relativity
generalized by Lovelock gravity to higher dimensions. The theory
also possesses exact solutions describing black holes
\cite{BHinLovelock}, whose thermodynamic behavior resembles the
one of Einstein-Hilbert black holes with a
modified entropy that is no longer proportional to the horizon's area \cite%
{Myers-Simon}.

Further physical input is in general required to select sensible theories
among Lovelock gravities (\ref{Lovelock}). For instance, a series of
inequivalent gravity actions has been presented in \cite%
{Crisostomo-Troncoso-Zanelli}, demanding the existence of a unique anti-de
Sitter (AdS) vacuum. In particular, Chern-Simons-AdS gravity in odd
dimensions \cite{Chamseddine} and Born-Infeld-AdS gravity in even dimensions
--often collectively referred to as {\em Dimensionally Continued Gravity}
\cite{Banados-Teitelboim-Zanelli-continuedBH}--, feature a symmetry
enhancement from local Lorentz to AdS group, that leaves the gravitational
constant $\kappa $ and the AdS radius $\ell $ as the only free parameters in
the theory.

As in standard gravity, Lovelock action with cosmological constant is
divergent in the infrared region and needs to be regularized. In the AdS/CFT
approach \cite{AdS/CFT} to the regularization problem, the finiteness of
Einstein-Hilbert action is achieved by the procedure known as holographic
renormalization \cite%
{Henningson-Skenderis,deHaro-Skenderis-Solodukhin,Skenderis-LectureNotes,Cai-Ohta}.
For a fixed boundary data $g_{(0)ij}$, this algorithm reconstructs
the spacetime metric solving iteratively the field equations in
the
Fefferman-Graham frame \cite{Fefferman-Graham}%
\begin{equation}
ds^{2}=G_{\mu \nu }\,dx^{\mu }dx^{\nu }=\frac{\ell ^{2}}{4\rho ^{2}}\,d\rho
^{2}+\frac{1}{\rho }\,g_{ij}(x,\rho )\,dx^{i}dx^{j}\,.  \label{radial}
\end{equation}%
Here, $g_{ij}(x,\rho )$\ is regular at the conformal boundary $\rho =0$,\ so
that it can be expanded in its vicinity as%
\begin{equation}
g_{ij}(x,\rho )=g_{(0)ij}(x)+\rho \,g_{(1)ij}(x)+\rho
^{2}g_{(2)ij}(x)+\cdots \,.  \label{gFG}
\end{equation}

This method results in the addition of boundary terms ${\cal
L}_{ct}$ to the bulk action (supplemented by the Gibbons-Hawking
term \cite{Gibbons-Hawking}), that are local functionals of the
boundary metric $h_{ij}=g_{ij}/\rho $, the intrinsic curvature
$R_{kl}^{ij}(h)$ and its covariant derivative $\nabla
_{m}R_{kl}^{ij}$. This construction is known as Dirichlet
counterterms procedure, what achieves a regularized action
\cite{Balasubramanian-Kraus,Emparan-Johnson-Myers}
\begin{equation}
I_{reg}=-\frac{1}{16\pi G}\int\limits_{M}d^{d+1}x\,\sqrt{-G}\left( \hat{R}%
-2\Lambda \right) -\frac{1}{8\pi G}\int\limits_{\partial M}d^{d}x\,\sqrt{-h}%
\,K+\int\limits_{\partial M}d^{d}x\,{\cal L}_{ct}(h,R(h),\nabla
R(h)). \label{IEHct}
\end{equation}%
In the above formula, $K$ is the trace of the extrinsic curvature.

However, the intrinsic regularization defined by this method
becomes technically involved in higher dimensions because of the
forbidding complexity of the equations for the coefficients
$g_{(k)}$ ($1\leq k\leq \left[ \frac{d}{2}\right] $) and the
plethora of possible covariant counterterms one could construct on
the boundary.

For higher-curvature theories, holographic renormalization
procedure would be even more cumbersome due to the highly
non-linear behavior of the equations of motion. In fact, the
regularization of quadratic curvature gravities has been carried
out only in particular cases by adding covariant local
counterterms that are not necessarily dictated by the holographic
renormalization procedure \cite{Nojiri-Odintsov-Ogushi}. For
Einstein-Gauss-Bonnet AdS gravity (the particular quadratic
combination of the curvature given by $p=2$ in
Eq.(\ref{Lovelock})), this approach provides the answer only for
the five-dimensional case \cite{Cvetic-Nojiri-Odintsov}. Thus, it
still leaves the open question on the form of the counterterms in
higher-dimensional Einstein-Gauss-Bonnet AdS, let alone in a
generic Lovelock gravity. Furthermore, in Dimensionally Continued
Gravity, the AdS vacuum is a zero of maximal degree in the field
equations, such that the first non-trivial relation for the
coefficients $g_{(k)}$ in (\ref{gFG}) will just appear at much
higher order in $\rho $ than the linear one.

Whichever the explicit form of the counterterms ${\cal L}_{ct}$ may be for
Lovelock-AdS gravity, the action (\ref{IEHct}) has to be promoted to the form%
\begin{equation}
I_{reg}=I+\kappa \int\limits_{\partial M}d^{d}x\,\beta
_{d}+\int\limits_{\partial M}d^{d}x\,{\cal L}_{ct}(h,R(h),\nabla
R(h))\,, \label{IregLovelock}
\end{equation}
such that the generalized Gibbons-Hawking term $\beta _{d}$
defines a variational principle for a Dirichlet boundary condition
on the metric for the action $I$ in Eq.(\ref{Lovelock}), what is
left unchanged by the addition of intrinsic counterterms. As we
will shown in detail below, the on-shell variation of the first
two terms in Eq.(\ref{IregLovelock}) adopts the canonical form
$\delta I=\int_{\partial M}d^{d}x\, \pi^{ij}\delta h_{ij}$, where
$\pi^{ij}$ corresponds to the momenta in a radial Hamiltonian
formulation for Lovelock gravity. Therefore, the role of the
counterterms ${\cal L}_{ct}$ is cancelling the divergences in the
canonical momenta, but it also means that the series should be
obtained from the integration of the divergent part of the
Hamiltonian variation in any gravity theory. This has been proved
in Ref.\cite{Papadimitriou-Skenderis}, and allowed to recover the
counterterm series in the Einstein-Hilbert case from the action of
the dilatations on the gravity fields expansion. Such strategy
might also be applied to higher curvature theories but, in
practice, such procedure for Lovelock gravity could be much more
complicated.

In view of the above arguments, it is quite remarkable that a universal
regularization prescription for any Lovelock theory with AdS asymptotics can
be provided using boundary terms with explicit dependence on the extrinsic
curvature $K_{ij}$, also known as Kounterterms series \cite%
{Kofinas-Olea-Lovelock}
\begin{equation}
{\cal I}_{reg}=I+c_{d}\int\limits_{\partial
M}d^{d}x\,B_{d}(h,R(h),K)\,. \label{IregKT}
\end{equation}%
Due to a profound connection to topological invariants (Euler term) and
Chern-Simons forms, the explicit form of this series only distinguishes even
from odd dimensions. The construction of the boundary terms $B_{d}$ does not
make use of the expansion in the metric (\ref{gFG}). Therefore, for a given
dimension, the Kounterterms expression remains the same regardless the
particular Lovelock gravity considered, even for Einstein-Hilbert \cite%
{OleaJHEP,OleaKounter} and Einstein-Gauss-Bonnet theories \cite{Kofinas-Olea}%
. Only the value of the coupling constant $c_{d}$\ is consistently tuned to
achieve a well-posed action principle in a given Lovelock-AdS theory.

The agreement between the proposal defined by Eq.(\ref{IregKT}) with the
standard regularization method, has been found --when the latter exists at
all-- at the level of the conserved quantities and Euclidean action for
asymptotically AdS (AAdS) solutions. In Einstein-Hilbert gravity, a direct
comparison between both procedures has been worked out in $2+1$ dimensions,
showing that the corresponding boundary prescriptions differs at most by a
topological invariant \cite{Miskovic-Olea}. For higher dimensions,
attempting a similar strategy would be in general very involved and not
particularly enlightening.

On the other hand, one might expect that further insight on this problem
would come out from other Lovelock theories, especially in view of the fact
that the form of $B_{d}$ is universal. But, unfortunately, in many cases
there is no even a counterterms series ${\cal L}_{ct}$ to compare with.

In this paper, we use Dimensionally Continued Gravity as a toy
model to perform the first explicit comparison between the
intrinsic and extrinsic regularization schemes in all dimensions.
This is only due to the fact that, in this theory, the obtention
of the Dirichlet counterterms from direct integration of the
divergent terms in the variation of the action is remarkably
simpler than in any other gravity theory.

This article is organized as follows. In the next section, we consider the
Dirichlet problem for an arbitrary Lovelock gravity, where the addition of a
generalized Gibbons-Hawking term defines a well-posed variational principle
for a Dirichlet boundary condition on the metric. This procedure is shown to
reproduce the Hamiltonian variation of the action for a radial foliation of
the spacetime. In Section \ref{Dirichlet}, for Dimensionally Continued
Gravity, the series ${\cal L}_{ct}$\ is obtained as a total variation of
local terms in the Dirichlet problem of the action. In Section \ref%
{Kounterterms}, we briefly review the Kounterterms construction for
Lovelock-AdS, specialized for Dimensionally Continued Gravity. Finally, we
show that the Dirichlet counterterms are generated simply taking the
difference between the Kounterterms series $c_{d}B_{d}$ and the generalized
Gibbons-Hawking term $\kappa \beta _{d}$.

\section{Dirichlet problem in Lovelock gravity \label{Dirichlet problem}}

In general, a well-defined action principle for gravity considers
supplementing the bulk Lagrangian by appropriate boundary terms such that
the on-shell action is stationary. This means that the surface terms coming
from an arbitrary variation of the action must be cancelled by choosing
suitable boundary conditions.

The Dirichlet problem for gravity consists in setting a well-posed action
principle by imposing a Dirichlet boundary condition on the metric. For
Einstein-Hilbert case, this is achieved by adding the Gibbons-Hawking
boundary term \cite{Gibbons-Hawking} to the bulk action. The systematic
construction of boundary terms that defines the Dirichlet problem in
Lovelock gravity was carried out by Myers in Ref.\cite{Myers}.

Let us briefly review this formalism. The Einstein-Hilbert term (that
corresponds to $p=1$ in (\ref{p-Lovelock})),
\begin{equation}
I^{(1)}=\int\limits_{M_{D}}\varepsilon _{A_{1}\cdots A_{D}}\, \hat{R}%
^{A_{1}A_{2}}e^{A_{3}}\cdots e^{A_{D}}\,,
\end{equation}
can be written as the dimensional continuation of the 2-dimensional Euler
term ${\cal E}_{2}=\varepsilon _{AB}\,\hat{R}^{AB}$, which is a topological
invariant. The variation of $I^{(1)}$ contributes to the equations of motion
and produces a surface term
\begin{equation}
\delta I^{(1)}=\int\limits_{\partial M_{D}}\varepsilon _{A_{1}\cdots
A_{D}}\,\delta \omega ^{A_{1}A_{2}}\,e^{A_{3}}\cdots e^{A_{D}}\,.
\label{var_I(1)}
\end{equation}
In the vicinity of the boundary, we take Gaussian (normal) coordinates
\begin{equation}
ds^{2}=G_{\mu \nu }\,dx^{\mu }dx^{\nu }=N^{2}(\rho )\,d\rho ^{2}+h_{ij}(\rho
,x)\,dx^{i}dx^{j}\,,  \label{Gaussian}
\end{equation}
and the corresponding local orthonormal frame
\begin{equation}
e^{1}=Nd\rho \,,\qquad e^{a}=e_{i}^{a}\,dx^{i}\,,
\end{equation}
with a splitting of the indices $A=\left( 1,a\right) $ for the tangent space
and $\mu =\left( \rho ,i\right) $ for the spacetime. When torsion vanishes,
the spin connection on $\partial M_{D}$ is
\begin{equation}
\omega ^{1a}=K^{a}=K_{i}^{j}\,e_{j}^{a}\,dx^{i},\qquad \omega ^{ab}=\omega
_{i}^{ab}(e_{j}^{c})\,dx^{i}\,,
\end{equation}
where $K_{ij}$ is the extrinsic curvature, that in the frame (\ref{Gaussian}%
) becomes
\begin{equation}
K_{ij}=-\frac{1}{2N}\,\partial _{\rho }h_{ij}\,.
\end{equation}

In this coordinate system, the variation (\ref{var_I(1)}) adopts the form
\begin{equation}
\delta I^{(1)}=-2\int\limits_{\partial M_{D}}\varepsilon _{a_{1}\cdots
a_{d}}\,\delta K^{a_{1}}\,e^{a_{2}}\cdots e^{a_{d}}\,,
\label{var_I(1)decomp}
\end{equation}%
where the Levi-Civita tensor at the boundary is defined by $\varepsilon
_{1a_{1}\cdots a_{d}}=-\varepsilon _{a_{1}\cdots a_{d}}$. The above surface
term contains the variation of the extrinsic curvature that must be
eliminated in the Dirichlet problem.

On the other hand, the integration of ${\cal E}_{2}$ over a two-dimensional
manifold without boundary is proportional to the Euler characteristic $\chi
(M_{2})$. When a boundary is introduced, the Euler theorem assigns a
boundary correction given by
\begin{equation}
\int\limits_{M_{2}}\varepsilon _{AB}\,\hat{R}^{AB}=-4\pi \,\chi
(M_{2})+\int\limits_{\partial M_{2}}\varepsilon _{AB}\,\theta ^{AB}\,.
\label{theorem}
\end{equation}%
Here $\theta ^{AB}=\omega ^{AB}-\bar{\omega}^{AB}$ stands for the Second
Fundamental Form, i.e., the difference between the dynamic field and a
reference spin connection that recovers Lorentz covariance at the boundary.
It is common to take $\bar{\omega}^{AB}$ as the spin connection from a
product metric that matches the geometry at the boundary, such that
\begin{equation}
\theta ^{1a}=K^{a}\,,\qquad \theta ^{ab}=0\,,
\end{equation}%
i.e., only normal components of the Second Fundamental Form are
non-vanishing at the boundary \cite%
{Eguchi-Gilkey-Hanson,Nakahara,Choquet-Dewitt}. From the dynamical point of
view, variations of both sides of Eq.(\ref{theorem}) produce $\varepsilon
_{AB}\,\delta \omega ^{AB}$ at the boundary.

Thus, in order to cancel the term (\ref{var_I(1)}) (or equivalently (\ref%
{var_I(1)decomp})), we dimensionally continue the boundary term in Eq.(\ref%
{theorem}), and obtain the Gibbons-Hawking term
\begin{eqnarray}
d^{d}x\,\beta ^{(1)} &=&-\varepsilon _{A_{1}\cdots A_{D}}\,\theta
^{A_{1}A_{2}}e^{A_{3}}\cdots e^{A_{D}}  \nonumber \\
&=&-2\left( D-2\right) !\,d^{d}x\,\sqrt{-h}\,K\,.
\end{eqnarray}%
Indeed, the variation of $I_{Dir}^{(1)}=I^{(1)}+\int_{\partial
M_{D}}d^{d}x\beta ^{(1)},$
\begin{eqnarray}
\delta I_{Dir}^{(1)} &=&2\left( D-2\right) \int\limits_{M_{D}}\varepsilon
_{a_{1}\cdots a_{d}}\,\delta e^{a_{1}}K^{a_{2}}e^{a_{3}}\cdots e^{a_{d}} \\
&=&\left( D-2\right) !\int\limits_{\partial M_{D}}d^{d}x\,\sqrt{-h}%
\,(h^{-1}\delta h)_{i}^{j}\,\left( K_{j}^{i}-\delta _{j}^{i}\,K\right) \,,
\end{eqnarray}%
has a suitable form to impose the Dirichlet boundary condition on the metric
$h_{ij}$.

In dimensions $D\geq 5$, the Gauss-Bonnet term (the second order term in the
Lovelock series)
\begin{eqnarray}
I^{(2)} &=&\int\limits_{M_{D}}\varepsilon _{A_{1}\cdots A_{D}}\,\hat{R}%
^{A_{1}A_{2}}\hat{R}^{A_{3}A_{4}}e^{A_{5}}\cdots e^{A_{D}}  \nonumber \\
&=&-\left( D-4\right) !\int\limits_{M_{D}}d^{D}x\,\sqrt{-G}\,\left( \hat{R}%
_{\mu \nu \alpha \beta }\,\hat{R}^{\mu \nu \alpha \beta }-4\hat{R}_{\mu \nu
}\,\hat{R}^{\mu \nu }+\hat{R}^{2}\right) \,,
\end{eqnarray}%
contributes to the bulk dynamics. In order to set the Dirichlet problem for
this term, one has to consider the Euler theorem in four dimensions,
\begin{equation}
\int\limits_{M_{4}}\varepsilon _{ABCD}\,\hat{R}^{AB}\hat{R}^{CD}=2\left(
4\pi \right) ^{2}\chi (M_{4})+2\int\limits_{\partial M_{4}}\varepsilon
_{ABCD}\,\theta ^{AB}\left( R^{CD}+\frac{1}{3}\,(\theta ^{2})^{CD}\right) ,
\label{4Euler}
\end{equation}%
where $R^{ab}=\frac{1}{2}\,R_{kl}^{ij}(h)\,e_{i}^{a}e_{j}^{b}\,dx^{k}dx^{l}$
is the intrinsic curvature and $R^{1a}=0$. The dimensional continuation of
the second Chern form (i.e., the boundary correction to the Euler
characteristic in (\ref{4Euler})) is \cite{Myers,Mueller-Hoissen}
\begin{eqnarray}
d^{d}x\,\beta ^{(2)} &=&-2\varepsilon _{A_{1}\cdots A_{D}}\,\theta
^{A_{1}A_{2}}\left( R^{A_{3}A_{4}}+\frac{1}{3}\,(\theta
^{2})^{A_{3}A_{4}}\right) e^{A_{5}}\cdots e^{A_{D}}  \nonumber \\
&=&4\varepsilon _{a_{1}\cdots a_{d}}\,K^{a_{1}}\left( R^{a_{2}a_{3}}-\frac{1%
}{3}\,K^{a_{2}}K^{a_{3}}\right) e^{a_{4}}\cdots e^{a_{d}}  \nonumber \\
&=&-4\left( D-4\right) !\,d^{d}x\,\sqrt{-h}\,\delta _{\lbrack
i_{1}i_{2}i_{3}]}^{[j_{1}j_{2}j_{3}]}\,K_{j_{1}}^{i_{1}}\left( \frac{1}{2}%
\,R_{j_{2}j_{3}}^{i_{2}i_{3}}(h)-\frac{1}{3}%
\,K_{j_{2}}^{i_{2}}K_{j_{3}}^{i_{3}}\right) \,,
\end{eqnarray}%
such that the corresponding Dirichlet variation is
\begin{equation}
\delta I_{Dir}^{(2)}=-2\left( D-4\right) !\int\limits_{\partial
M_{D}}d^{d}x\,\sqrt{-h}\,\delta _{\lbrack
i\,i_{1}i_{2}i_{3}]}^{[j\,j_{1}j_{2}j_{3}]}\,(h^{-1}\delta
h)_{j}^{i}\,K_{j_{1}}^{i_{1}}\left( \frac{1}{2}%
\,R_{j_{2}j_{3}}^{i_{2}i_{3}}(h)-\frac{1}{3}%
\,K_{j_{2}}^{i_{2}}K_{j_{3}}^{i_{3}}\right) \,.  \label{var_GB}
\end{equation}%
We have used the Gauss-Codazzi relations at the boundary
\begin{eqnarray}
\hat{R}^{ab} &=&R^{ab}-K^{a}K^{b}\,, \\
\hat{R}^{1a} &=&DK^{a}\,,
\end{eqnarray}%
or equivalently
\begin{eqnarray}
\hat{R}_{kl}^{ij} &=&R_{kl}^{ij}(h)-K_{k}^{i}K_{l}^{j}+K_{l}^{i}K_{k}^{j}\,,
\label{GC1} \\
\hat{R}_{jk}^{i\rho } &=&\frac{1}{N}\left( \nabla _{j}K_{k}^{i}-\nabla
_{k}K_{j}^{i}\right) \,,  \label{GC2}
\end{eqnarray}%
where $D_{i}=D_{i}(\omega )$ and $\nabla _{i}=\nabla _{i}(\Gamma )$ are
covariant derivatives with respect to the spin connection and Christoffel
symbol, respectively.

For arbitrary $p$, the generalized Gibbons-Hawking term is
\begin{eqnarray}
d^{d}x\,\beta ^{(p)} &=&-p\int\limits_{0}^{1}dt\,\varepsilon _{A_{1}\cdots
A_{D}}\,\theta ^{A_{1}A_{2}}\left( R^{A_{3}A_{4}}+t^{2}(\theta
^{2})^{A_{3}A_{4}}\right) \times \cdots  \nonumber \\
&&\qquad \qquad \cdots \times \left( R^{A_{2p-1}A_{2p}}+t^{2}(\theta
^{2})^{A_{2p-1}A_{2p}}\right) e^{A_{2p+1}}\cdots e^{A_{D}} \\
&=&2p\int\limits_{0}^{1}dt\,\varepsilon _{a_{1}\cdots
a_{d}}\,K^{a_{1}}\left( R^{a_{2}a_{3}}-t^{2}K^{a_{2}}K^{a_{3}}\right) \times
\cdots  \nonumber \\
&&\qquad \qquad \cdots \times \left(
R^{a_{2p-2}a_{2p-1}}-t^{2}K^{a_{2p-2}}K^{a_{2p-1}}\right) e^{a_{2p}}\cdots
e^{a_{d}}\,,
\end{eqnarray}%
or in tensorial notation
\begin{eqnarray}
d^{d}x\,\beta ^{(p)} &=&-2p\left( D-2p\right)
!\,d^{d}x\,\int\limits_{0}^{1}dt\,\delta _{\lbrack i_{1}\cdots
i_{2p-1}]}^{[j_{1}\cdots j_{2p-1}]}\,K_{j_{1}}^{i_{1}}\left( \frac{1}{2}%
\,R_{j_{2}j_{3}}^{i_{2}i_{3}}(h)-t^{2}\,K_{j_{2}}^{i_{2}}K_{j_{3}}^{i_{3}}%
\right) \times \cdots  \nonumber \\
&&\qquad \qquad \cdots \times \left( \frac{1}{2}%
\,R_{j_{2p-2}j_{2p-1}}^{i_{2p-2}i_{2p-1}}(h)-t^{2}%
\,K_{j_{2p-2}}^{i_{2p-2}}K_{j_{2p-1}}^{i_{2p-1}}\right) \,.
\end{eqnarray}%
It is worthwhile noticing that the procedure of dimensional continuation of
a given Chern form to define the Dirichlet problem in Lovelock gravity does
not work in spacetimes with torsion (Riemann-Cartan theory).

The Dirichlet variation for the $p$-th term of Lovelock series takes the
form
\begin{eqnarray}
\delta I_{Dir}^{(p)} &=&-p\,\left( D-2p\right) !\int\limits_{\partial
M_{D}}d^{d}x\,\sqrt{-h}\,\int\limits_{0}^{1}dt\,\delta _{\lbrack
ii_{1}\cdots i_{2p-1}]}^{[jj_{1}\cdots j_{2p-1}]}\,(h^{-1}\delta
h)_{j}^{i}\,K_{j_{1}}^{i_{1}}\,\times  \nonumber \\
&& \times \left( \frac{1}{2}%
\,R_{j_{2}j_{3}}^{i_{2}i_{3}}(h)-t^{2}K_{j_{2}}^{i_{2}}K_{j_{3}}^{i_{3}}%
\right) \cdots \left( \frac{1}{2}%
\,R_{j_{2p-2}j_{2p-1}}^{i_{2p-2}i_{2p-1}}(h)-t^{2}%
\,K_{j_{2p-2}}^{i_{2p-2}}K_{j_{2p-1}}^{i_{2p-1}}\right) .
\end{eqnarray}%
As a consequence, the Lovelock action set for the Dirichlet problem is
\begin{equation}
I_{Dir}=I+\kappa \int\limits_{\partial M_{D}}d^{d}x\beta _{d}\,,
\label{L-Dirichlet}
\end{equation}%
where the boundary term is given by
\begin{equation}
\beta _{d}=\sum\limits_{p=0}^{[(D-1)/2]}\alpha _{p}\,\beta ^{(p)}\,.
\end{equation}%
Finally, the variation of the Dirichlet action can be written as
\begin{eqnarray}
\delta I_{Dir} &=&-\kappa \sum\limits_{p=0}^{[(D-1)/2]}\alpha
_{p}\,p\,\left( D-2p\right) !\int\limits_{\partial M_{D}}d^{d}x\,\sqrt{-h}%
\,\int\limits_{0}^{1}dt\,\delta _{\lbrack ii_{1}\cdots
i_{2p-1}]}^{[jj_{1}\cdots j_{2p-1}]}\,(h^{-1}\delta
h)_{j}^{i}\,K_{j_{1}}^{i_{1}}\,\times  \nonumber \\
&&\times \left( \frac{1}{2}%
\,R_{j_{2}j_{3}}^{i_{2}i_{3}}(h)-t^{2}K_{j_{2}}^{i_{2}}K_{j_{3}}^{i_{3}}%
\right) \cdots \left( \frac{1}{2}%
\,R_{j_{2p-2}j_{2p-1}}^{i_{2p-2}i_{2p-1}}(h)-t^{2}%
\,K_{j_{2p-2}}^{i_{2p-2}}K_{j_{2p-1}}^{i_{2p-1}}\right) .
\label{var_Dirichlet}
\end{eqnarray}%
The parametric integration can be performed explicitly, and using the
relation between spacetime and induced Riemann tensors (\ref{GC1}) produces
\begin{eqnarray*}
&&\int\limits_{0}^{1}dt\,\delta _{\lbrack ii_{1}\cdots
i_{2p-1}]}^{[jj_{1}\cdots j_{2p-1}]}\,K_{j_{1}}^{i_{1}}\left( \frac{1}{2}%
\,R_{j_{2}j_{3}}^{i_{2}i_{3}}(h)-t^{2}K_{j_{2}}^{i_{2}}K_{j_{3}}^{i_{3}}%
\right) \cdots \left( \frac{1}{2}%
\,R_{j_{2p-2}j_{2p-1}}^{i_{2p-2}i_{2p-1}}(h)-t^{2}%
\,K_{j_{2p-2}}^{i_{2p-2}}K_{j_{2p-1}}^{i_{2p-1}}\right) \\
&=&\frac{1}{2^{p+1}}\,\delta _{\lbrack ii_{1}\cdots
i_{2p-1}]}^{[jj_{1}\cdots j_{2p-1}]}\sum_{s=0}^{p-1}\frac{4^{p-s}\left(
p-1\right) !}{s!\left( 2p-2s-1\right) !!}\,\hat{R}_{j_{1}j_{2}}^{i_{1}i_{2}}%
\cdots \hat{R}_{j_{2s-1}j_{2s}}^{i_{2s-1}i_{2s}}\,K_{j_{2s+1}}^{i_{2s+1}}%
\cdots K_{i_{2p-1}}^{i_{2p-1}}\,.
\end{eqnarray*}%
It is clear from the last line that the Dirichlet variation agrees with the
variation of the action in the Hamiltonian formulation of Lovelock gravity
\cite{Teitelboim-Zanelli} for the\ radial foliation of spacetime considered
in \cite{Banados-Olea-Theisen},
\begin{equation}
\delta I_{H}=\int\limits_{\partial M_{D}}d^{d}x\,(h^{-1}\delta
h)_{j}^{i}\,\pi _{i}^{j}\,,  \label{Hamilton}
\end{equation}%
where the canonical momenta have the form
\begin{eqnarray}
\pi _{i}^{j} &=&-\kappa \sum_{p=1}^{[\left( D-1\right) /2]}\frac{\left(
D-2p\right) !\,p!}{2^{p+1}}\,\alpha _{p}\sum_{s=0}^{p-1}C_{s(p)}\,\left( \pi
_{s(p)}\right) _{i}^{j}\,, \\
\left( \pi _{s(p)}\right) _{i}^{j} &=&\sqrt{-h}\,\delta _{\lbrack
ii_{1}\cdots i_{2p-1}]}^{[jj_{1}\cdots j_{2p-1}]}\,\hat{R}%
_{j_{1}j_{2}}^{i_{1}i_{2}}\cdots \hat{R}_{j_{2s-1}j_{2s}}^{i_{2s-1}i_{2s}}%
\,K_{j_{2s+1}}^{i_{2s+1}}\cdots K_{i_{2p-1}}^{i_{2p-1}}\,,
\end{eqnarray}%
and the coefficients $C_{s(p)}$ are given by
\begin{equation}
C_{s(p)}=\frac{4^{p-s}}{s!\left( 2p-2s-1\right) !!}\,.
\end{equation}

In the Lagrangian formalism, the variation of the action with
respect to the metric defines a quasilocal (boundary) stress
tensor \cite{Brown-York}, that can be therefore identified with
the canonical momenta in Hamiltonian formalism. The above
relations are also useful to study the generalized Israel junction
conditions for branes in Lovelock gravity, as the discontinuity in
the canonical momenta, and where the boundary is the brane itself
\cite{Gravanis-WillisonJGeom} (for the Einstein-Gauss-Bonnet case,
see \cite{SCDavis,Gravanis-WillisonPLB}).

In Lovelock gravity with negative cosmological constant both the
action and the stress tensor (or, equivalently, the canonical
momenta) are in general divergent. Therefore, the regularization
problem requires the addition of local counterterms, such that
their inclusion does not spoil the action principle based on a
Dirichlet boundary condition on the metric. For a given
Lovelock-AdS theory, there is no a systematic way to generate the
counterterms series, and even in the EH case it not possible to
provide a closed form for ${\cal L}_{ct}$. However, as shown in
Ref.\cite{Papadimitriou-Skenderis}, assuming AdS asymptotics, the
divergent part of the radial canonical momenta is linked to the
divergent part of the on-shell Lagrangian. The Hamilton-Jacobi
relations imply that the normalizable modes of the fields
expansion do not produce additional divergences and thus, the
counterterms are always local. This argument opens the possibility
of obtaining the Dirichlet counterterms from direct integration of
the divergent parts of the Hamiltonian variation. This procedure
can be performed for Chern-Simons-AdS gravity which, on the
contrary to the Einstein-Hilbert case, produces a closed form for
the Dirichlet counterterms (and conformal anomaly) for all odd
dimensions \cite{Banados-Olea-Theisen}. We shall show below that
the same method can be carried out (using either Hamiltonian or
Lagrangian formulation) in another Lovelock theory
(Born-Infeld-AdS), which can be regarded the even-dimensional
counterpart of Chern-Simons-AdS, because global AdS spacetime is
also a solution of maximal rank in the equations of motion.

\section{Dirichlet counterterms \label{Dirichlet}}

Let us briefly review the construction of Dirichlet counterterms for
Chern-Simons-AdS gravity discussed in \cite{Banados-Olea-Theisen}.

\subsection{Chern-Simons-AdS}

Chern-Simons gravity was first considered in \cite{Chamseddine} in
five dimensions and in higher odd dimensions in
\cite{Chamseddine-super,Troncoso-Zanelli}.

Unlike in three dimensions, higher-dimensional Chern-Simons gravity is not
topological, but possesses propagating degrees of freedom \cite%
{Banados-Garay-Henneaux} which number may vary from a sector to another in
the phase space \cite{Miskovic-Zanelli}. When the number of degrees of
freedom is fewer than maximal, it is said that the sector is irregular \cite%
{Miskovic-Zanelli}. The AdS space in pure Chern-Simons gravity is an example
of such an irregular solution, and in its vicinity gravity becomes
topological. However, the AdS vacuum can also be stable, as it was shown in
five-dimensional Chern-Simons-AdS supergravity \cite%
{Miskovic-Troncoso-Zanelli}.

In Chern-Simons-AdS gravity, the Lagrangian comes from a Chern-Simons
density for the group $SO(2n,2)$ in $D=2n+1$ dimensions, and corresponds to
the particular choice of the coefficients $\alpha _{p}$%
\begin{equation}
\alpha _{p}:=\frac{\ell ^{2(p-n)}}{D-2p}\left(
\begin{array}{c}
n \\
p%
\end{array}%
\right) \,,\qquad p\leq n\,,  \label{CScoeff}
\end{equation}%
that allows the action to be rewritten as an integration over the continuous
parameter $t$\thinspace ,
\begin{eqnarray}
I_{2n+1} &=&\kappa
\int\limits_{M_{2n+1}}\int\limits_{0}^{1}dt\,\varepsilon
_{A_{1}\cdots A_{2n+1}}\left(
\hat{R}^{A_{1}A_{2}}+\frac{t^{2}}{\ell ^{2}}
\,e^{A_{1}}e^{A_{2}}\right) \times   \nonumber \\
&&\qquad \qquad \qquad \cdots \times \left(
\hat{R}^{A_{2n-1}A_{2n}}+\frac{ t^{2}}{\ell
^{2}}\,e^{A_{2n-1}}e^{A_{2n}}\right) e^{A_{2n+1}}\,, \label{CS}
\end{eqnarray}
The field equations following from the above action are
\begin{equation}
E_{\nu }^{\mu }=\delta _{\lbrack \nu \nu _{1}\cdots \nu _{2n}]}^{[\mu \mu
_{1}\cdots \mu _{2n}]}\,\left( \hat{R}_{\mu _{1}\mu _{2}}^{\nu _{1}\nu _{2}}+%
\frac{1}{\ell ^{2}}\,\delta _{\lbrack \mu _{1}\mu _{2}]}^{[\nu _{1}\nu
_{2}]}\right) \cdots \left( \hat{R}_{\mu _{2n-1}\mu _{2n}}^{\nu _{2n-1}\nu
_{2n}}+\frac{1}{\ell ^{2}}\,\delta _{\lbrack \mu _{2n-1}\mu _{2n}]}^{[\nu
_{2n-1}\nu _{2n}]}\right) =0\,.  \label{CS-eom}
\end{equation}%
From now on, we set the AdS radius $\ell =1$.

In order to cast the variation of the action into the form (\ref{Hamilton}),
we supplement the bulk action with the corresponding Gibbons-Hawking-Myers
term
\begin{eqnarray}
\beta _{2n} &=&-2n\,\sqrt{-h}\int\limits_{0}^{1}dt\,\int\limits_{0}^{1}ds\,%
\delta _{\left[ j_{1}...j_{2n-1}\right] }^{\left[ i_{1}...i_{2n-1}\right]
}\,K_{i_{1}}^{j_{1}}\left( {\frac{1}{2}\,}%
R_{i_{2}i_{3}}^{j_{2}j_{3}}(h)-t^{2}K_{i_{2}}^{j_{2}}K_{i_{3}}^{j_{3}}+s^{2}%
\delta _{i_{2}}^{j_{2}}\delta _{i_{3}}^{j_{3}}\right) \times \cdots
\nonumber \\
&&\cdots \times \left( {\frac{1}{2}\,}%
R_{i_{2n-2}i_{2n-1}}^{j_{2n-2}j_{2n-1}}(h)-t^{2}K_{i_{2n-2}}^{j_{2n-2}}K_{i_{2n-1}}^{j_{2n-1}}+s^{2}\delta _{i_{2n-2}}^{j_{2n-2}}\delta _{i_{2n-1}}^{j_{2n-1}}\right) .
\label{GH-CS}
\end{eqnarray}%
Therefore, the variation of the action for the Dirichlet problem $%
I_{Dir}=I_{2n+1}+\kappa \int_{\partial M_{2n+1}}d^{2n}x\beta _{2n}$ is given
by the expression
\begin{eqnarray}
\delta I_{2n+1}^{Dir} &=&-n\kappa \int\limits_{\partial M_{2n+1}}d^{2n}x\,%
\sqrt{-h}\int\limits_{0}^{1}dt\,\delta _{\left[ jj_{1}\cdots j_{2n-1}\right]
}^{\left[ ii_{1}\cdots i_{2n-1}\right] }\,(h^{-1}\delta
h)_{i}^{j}\,K_{i_{1}}^{j_{1}}\left( {\frac{1}{2}\,}%
R_{i_{2}i_{3}}^{j_{2}j_{3}}(h)-t^{2}K_{i_{2}}^{j_{2}}K_{i_{3}}^{j_{3}}+%
\delta _{i_{2}}^{j_{2}}\delta _{i_{3}}^{j_{3}}\right) \times
\nonumber \\
&&\cdots \times \left( {\frac{1}{2}\,}%
R_{i_{2n-2}i_{2n-1}}^{j_{2n-2}j_{2n-1}}(h)-t^{2}K_{i_{2n-2}}^{j_{2n-2}}K_{i_{2n-1}}^{j_{2n-1}}+\delta _{i_{2n-2}}^{j_{2n-2}}\delta _{i_{2n-1}}^{j_{2n-1}}\right) \,.
\label{var_D-CS}
\end{eqnarray}%
As we had mentioned above, this variation also defines the quasilocal stress
tensor $T_{j}^{i}(h)\,$. In order to identify divergences and finite part of
this expression, we use the Fefferman-Graham form of the metric
\begin{eqnarray}
h_{ij} &=&\frac{1}{\rho }\,g_{ij}\,,  \label{rescaled h} \\
K_{j}^{i} &=&\delta _{j}^{i}-\rho \,k_{j}^{i}\,,  \label{k}
\end{eqnarray}%
where the rescaled metric $g_{ij}$ is given in (\ref{gFG}) and $%
k_{j}^{i}=g^{ik}\partial _{\rho }g_{kj}$ are regular at the conformal
boundary. Any AAdS metric can be brought into this form near $\rho =0$.
Other useful relations are
\begin{eqnarray}
R_{kl}^{ij}(h) &=&\rho \,R_{kl}^{ij}(g)\,,  \label{rescaled R} \\
\sqrt{-h} &=&\frac{\sqrt{-g}}{\rho ^{n}}\,,  \label{det h} \\
(h^{-1}\delta h)_{i}^{j} &=&(g^{-1}\delta g)_{i}^{j}.  \label{rescaled var}
\end{eqnarray}

It can be shown that, on the boundary, the divergent terms in (\ref{var_D-CS}%
) do not depend on $k_{j}^{i}$,
\begin{eqnarray}
\delta I_{2n+1}^{Dir} &=&-n!\kappa \int\limits_{\partial M_{2n+1}}d^{2n}x\,%
\sqrt{-g}\,\sum\limits_{p=0}^{n-1}\frac{\left( n-p\right) !2^{2n-3p-2}}{p!}\,%
\frac{1}{\rho ^{n-p}}\times  \nonumber \\
&&\qquad \qquad \times \,\delta _{\left[ jj_{1}\cdots j_{2p}\right] }^{\left[
ii_{1}\cdots i_{2p}\right] }\,(g^{-1}\delta
g)_{i}^{j}\,R_{i_{1}i_{2}}^{j_{1}j_{2}}(g)\cdots
R_{i_{2p-2}i_{2p}}^{j_{2p-2}j_{2p}}(g)+{\cal O}(1)\,,  \label{var_CS_div}
\end{eqnarray}%
so that they can be integrated out as local functions of the boundary metric
$h_{ij}$. These terms must be put back into the original action, with the
opposite sign, playing the role of Dirichlet counterterms ${\cal L}_{2n}$,
\begin{equation}
{\cal L}_{2n}=n!\kappa \,\sqrt{-h}\,\sum\limits_{p=0}^{n-1}\frac{%
2^{2n-3p-1}\left( n-p-1\right) !}{p!}\,\,\delta _{\left[ j_{1}\cdots j_{2p}%
\right] }^{\left[ i_{1}\cdots i_{2p}\right] }%
\,R_{i_{1}i_{2}}^{j_{1}j_{2}}(h)\cdots
R_{i_{2p-2}i_{2p}}^{j_{2p-2}j_{2p}}(h)\,,  \label{ct-CS}
\end{equation}%
such that the total action%
\begin{equation}
I_{2n+1}^{reg}=I_{2n+1}^{Dir}+\int\limits_{\partial M_{2n+1}}d^{2n}x\,{\cal L%
}_{2n}\,
\end{equation}%
is regularized.

The finite part in the Dirichlet variation (\ref{var_CS_div}) when $\rho
\rightarrow 0$ produces a regularized stress tensor,
\begin{equation}
T_{j}^{i}=\frac{2}{\sqrt{-g_{(0)}}}\,g_{(0)jk}\,\frac{\delta I_{2n+1}^{reg}}{%
\delta g_{(0)ki}}\,,  \label{T-reg}
\end{equation}%
which is related to the quasilocal stress tensor $T_{j}^{i}(h)$ as
\begin{equation}
T_{j}^{i}=\lim_{\rho \rightarrow 0}\frac{1}{\rho ^{\frac{d}{2}}}%
\,T_{j}^{i}(h)\,,
\end{equation}%
and takes the form
\begin{equation}
T_{j}^{i}=2n\kappa \,\int\limits_{0}^{1}dt\,\delta _{\left[ jj_{1}\cdots
j_{2n-1}\right] }^{\left[ ii_{1}\cdots i_{2n-1}\right] }\,k_{i_{1}}^{j_{1}}%
\left( {\frac{1}{2}\,}R_{i_{2}i_{3}}^{j_{2}j_{3}}(g)+2t\,k_{i_{2}}^{j_{2}}%
\delta _{i_{3}}^{j_{3}}\right) \cdots \left( {\frac{1}{2}\,}%
R_{i_{2n-2}i_{2n-1}}^{j_{2n-2}j_{2n-1}}(g)+2t\,k_{i_{2n-2}}^{j_{2n-2}}\delta
_{i_{2n-1}}^{j_{2n-1}}\right) \,.
\end{equation}%
The trace of the above stress tensor leads to a Weyl anomaly proportional to
the Euler density in any $d=2n$ dimension (type A) \cite%
{Banados-Olea-Theisen,Banados-Schwimmer-Theisen}. A regularization mechanism
for five-dimensional Chern-Simons-AdS gravity with Dirichlet boundary
conditions, that defines a stress tensor in Riemann-Cartan spacetimes was
considered in Ref.\cite{Banados-Miskovic-Theisen}.

In an arbitrary Lovelock gravity, the procedure of integrating out
the divergent pieces as local counterterms would be more intricate
because, in general, the power series in $\rho $ would contain
$k_{j}^{i}$, and it would be necessary to prove explicitly that
there are no non-local contributions. In the Chern-Simons-AdS
case, the symmetry enhancement of the theory seems to be
responsible for the simple obtention of the counterterms series.

%%%%%%%%%%%%%%%%%%%%%%%%%%%%%%%%%%%%%%%%%%%%%%%%%%%%%%%%%%%%%%%%%%%%%%%%%%%%%%%%%%%

\subsection{Born-Infeld-AdS}

Born-Infeld gravity in even dimensions ($D=2n$) corresponds to the
coefficients set
\begin{equation}
\alpha _{p}:=\ell ^{2(p-n)}\left(
\begin{array}{c}
n \\
p%
\end{array}
\right) \,,\qquad p\leq n-1\,,
\end{equation}
that allows the action to be written in the form
\begin{eqnarray}
I_{2n} &=&n\kappa
\int\limits_{M_{2n}}\int\limits_{0}^{1}du\,\varepsilon
_{A_{1}\cdots A_{2n}}\,\left(
\hat{R}^{A_{1}A_{2}}+u\,e^{A_{1}}e^{A_{2}}
\right) \times   \nonumber \\
&&\qquad \qquad \cdots \times \left( \hat{R}^{A_{2n-1}A_{2n-2}}+u
\,e^{A_{2n-1}}e^{A_{2n-2}}\right) e^{A_{2n-1}}e^{A_{2n}}\,,
\label{BI-bulk}
\end{eqnarray}
using the identity (\ref{binomial}) from Appendix A. The equations
of motion derived from this action are
\begin{equation}
E_{\nu }^{\mu }=\delta _{\lbrack \nu \nu _{1}\cdots \nu _{2n-2}]}^{[\mu \mu
_{1}\cdots \mu _{2n-2}]}\,\left( \hat{R}_{\mu _{1}\mu _{2}}^{\nu _{1}\nu
_{2}}+\delta _{\lbrack \mu _{1}\mu _{2}]}^{[\nu _{1}\nu _{2}]}\right) \cdots
\left( \hat{R}_{\mu _{2n-3}\mu _{2n}-2}^{\nu _{2n-3}\nu _{2n-2}}+\delta
_{\lbrack \mu _{2n-3}\mu _{2n-2}]}^{[\nu _{2n-3}\nu _{2n-2}]}\right) =0\,.
\label{EOMBI}
\end{equation}%
The generalized Gibbons-Hawking term in this case is
\begin{eqnarray}
\beta _{2n-1} &=&-4n\left( n-1\right) \,\sqrt{-h}\int\limits_{0}^{1}dt\,\int%
\limits_{0}^{1}ds\,\delta _{\left[ j_{1}...j_{2n-3}\right] }^{\left[
i_{1}...i_{2n-3}\right] }\,K_{i_{1}}^{j_{1}}\left( {\frac{1}{2}\,}%
R_{i_{2}i_{3}}^{j_{2}j_{3}}(h)-t^{2}K_{i_{2}}^{j_{2}}K_{i_{3}}^{j_{3}}+s\,%
\delta _{i_{2}}^{j_{2}}\delta _{i_{3}}^{j_{3}}\right) \times
\nonumber \\
&&\!\!\!\!\cdots \times \left( {\frac{1}{2}\,}%
R_{i_{2n-4}i_{2n-3}}^{j_{2n-4}j_{2n-3}}(h)-t^{2}K_{i_{2n-4}}^{j_{2n-4}}K_{i_{2n-3}}^{j_{2n-3}}+s\,\delta _{i_{2n-4}}^{j_{2n-4}}\delta _{i_{2n-3}}^{j_{2n-3}}\right) \,,
\end{eqnarray}%
and the variation of the Dirichlet action
$I_{2n}^{Dir}=I_{2n}+\kappa \int_{\partial M_{2n}}d^{2n-1}x\,\beta
_{2n-1}$ is given by the expression
\begin{eqnarray} \delta
I_{2n}^{Dir} &=&-2n\left( n-1\right) \kappa \int\limits_{\partial
M_{2n}}d^{2n-1}x\,\sqrt{-h}\int\limits_{0}^{1}dt\,\delta _{\left[
jj_{1}\cdots j_{2n-3}\right] }^{\left[ ii_{1}\cdots
i_{2n-3}\right]
}\,(h^{-1}\delta h)_{i}^{j}\,K_{i_{1}}^{j_{1}}\times   \nonumber \\
&&\qquad \times \left( {\frac{1}{2}\,}%
R_{i_{2}i_{3}}^{j_{2}j_{3}}(h)-t^{2}K_{i_{2}}^{j_{2}}K_{i_{3}}^{j_{3}}+
\delta _{i_{2}}^{j_{2}}\delta _{i_{3}}^{j_{3}}\right) \times
\cdots
\nonumber \\
&&\qquad \qquad \cdots \times \left( {\frac{1}{2}\,}%
R_{i_{2n-4}i_{2n-3}}^{j_{2n-4}j_{2n-3}}(h)-t^{2}K_{i_{2n-4}}^{j_{2n-4}}K_{i_{2n-3}}^{j_{2n-3}}+\delta
_{i_{2n-4}}^{j_{2n-4}}\delta _{i_{2n-3}}^{j_{2n-3}}\right) \,.
\label{var_D-BI}
\end{eqnarray}
Using the Fefferman-Graham form of the metric, in the limit $\rho
\rightarrow 0,$ we find that the divergent terms in $\delta
I_{2n}^{Dir}$ do not depend on $k_{j}^{i}$ until order $\rho
^{-3/2}$,
\begin{eqnarray}
\delta I_{2n}^{Dir} &=&-n!\,\kappa \int\limits_{\partial M_{2n}}d^{2n-1}x\,%
\sqrt{-g}\,\sum\limits_{p=0}^{n-2}\frac{2^{2n-3p-2}\left( n-p-1\right) !}{p!}%
\,\frac{1}{\rho ^{n-p-\frac{1}{2}}}\times  \nonumber \\
&&\qquad \qquad \times \,\delta _{\left[ jj_{1}\cdots j_{2p}\right] }^{\left[
ii_{1}\cdots i_{2p}\right] }\,(g^{-1}\delta
g)_{i}^{j}\,R_{i_{1}i_{2}}^{j_{1}j_{2}}(g)\cdots
R_{i_{2p-1}i_{2p}}^{j_{2p-1}j_{2p}}(g)+{\cal O}(\rho ^{-1/2})\,.
\end{eqnarray}%
Again, they can be integrated out as local functions of the boundary metric
\begin{equation}
{\cal L}_{2n-1}=n!\,\kappa \,\sqrt{-h}\,\sum\limits_{p=0}^{n-2}\frac{%
2^{2n-3p-1}\left( n-p-1\right) !}{p!}\,\delta _{\left[ j_{1}\cdots j_{2p}%
\right] }^{\left[ i_{1}\cdots i_{2p}\right] }%
\,R_{i_{1}i_{2}}^{j_{1}j_{2}}(h)\cdots
R_{i_{2p-1}i_{2p}}^{j_{2p-1}j_{2p}}(h)\,,  \label{CT-BI}
\end{equation}%
and should be added to the original Dirichlet action as divergent
counterterms
\begin{equation}
I_{2n}^{reg}=I_{2n}^{Dir}+\int\limits_{\partial M_{2n}}d^{2n-1}x\,{\cal L}%
_{2n-1}\,.
\end{equation}

What is left at the boundary, after the regularization with the Dirichlet
counterterms (\ref{CT-BI})
\begin{eqnarray}
\delta I_{2n}^{reg} &=&\frac{2n\left( n-1\right) \kappa }{\sqrt{\rho }}%
\int\limits_{\partial M_{2n}}d^{2n-1}x\,\sqrt{-g}\int\limits_{0}^{1}dt\,%
\delta _{\left[ jj_{1}\cdots j_{2n-3}\right] }^{\left[ ii_{1}\cdots i_{2n-3}%
\right] }\,(g^{-1}\delta g)_{i}^{j}\,k_{i_{1}}^{j_{1}}\times  \nonumber \\
&&\qquad \times \left( {\frac{1}{2}\,}R_{i_{2}i_{3}}^{j_{2}j_{3}}+2t%
\,k_{i_{2}}^{j_{2}}\delta _{i_{3}}^{j_{3}}\right) \cdots \left( {\frac{1}{2}%
\,}R_{i_{2n-4}i_{2n-3}}^{j_{2n-4}j_{2n-3}}+2t\,k_{i_{2n-4}}^{j_{2n-4}}\delta
_{i_{2n-3}}^{j_{2n-3}}\right)  \label{TDirBI}
\end{eqnarray}%
defines a finite stress tensor for Born-Infeld-AdS gravity, that does not
provide the correct conserved quantities for static black hole solutions
found in \cite{Banados-Teitelboim-Zanelli-continuedBH}. In the corresponding
section below, it is shown that the difference respect a stress tensor
obtained from the extrinsic regularization of the action (\ref{IregKT}) is
at most a finite contribution.

%%%%%%%%%%%%%%%%%%%%%%%%%%%%%%%%%%%%%%%%%%%%%%%%%%%%%%%%%%%%%%%%%%%%%%%%%%%%%%%%%%%

\section{Kounterterms \label{Kounterterms}}

In the standard Dirichlet formulation of AdS gravity, the counterterms
introduced to regularize the action are covariant functionals of the
boundary metric, the intrinsic curvature and covariant derivatives of the
intrinsic curvature. When varied, they preserve a Dirichlet boundary
condition for the metric.

On the other hand, it has been recently introduced an alternative
regularization procedure that consists in addition of boundary terms that
contain explicit dependence on the extrinsic curvature $K_{ij}$
(Kounterterms). This choice necessarily modifies the boundary conditions
required to attain a well-posed action principle. In particular, the surface
term coming from the on-shell variation of the action will contain
variations of the extrinsic curvature that are usually cancelled by a
generalized Gibbons-Hawking term in the Dirichlet formulation of gravity.

%%%%%%%%%%%%%%%%%%%%%%%%%%%%%%%%%%%%%%%%%%%%%%%%%%%%%%%%%%%%%%%%%%%%%%%%%%%%%%%%%%%

\subsection{Chern-Simons-AdS}

A boundary term that regularizes the action for Chern-Simons-AdS gravity was
constructed in Ref.\cite{Mora-Olea-Troncoso-Zanelli-CS}, based on a
well-posed action principle where the extrinsic curvature is kept fixed at
the boundary. It was further clarified in \cite{Miskovic-Olea} that this
boundary condition arises naturally from the asymptotic form of the fields
in Fefferman-Graham expansion. As a consequence, this condition is suitable
to treat the variational problem in a large set of gravity theories that
support AAdS solutions. The corresponding boundary term $B_{2n}$\ that
regulates the conserved quantities and Euclidean action in Chern-Simons-AdS
gravity, provides also the correct answer for Einstein-Hilbert case \cite%
{OleaKounter,Mora-Olea-Troncoso-Zanelli-odd}, Einstein-Gauss-Bonnet gravity
\cite{Kofinas-Olea} and a generic Lovelock-AdS theory \cite%
{Kofinas-Olea-Lovelock}.

We consider the Chern-Simons-AdS action in $2n+1$ dimensions,
\begin{equation}
{\cal I}_{2n+1}=I_{2n+1}+c_{2n}\int\limits_{\partial
M_{2n+1}}d^{2n}x\,B_{2n}\,,  \label{I2n+1BC}
\end{equation}%
supplemented by a boundary term $B_{2n}$,
\begin{eqnarray}
B_{2n} &=&-n\int\limits_{0}^{1}dt\int\limits_{0}^{t}ds\,\varepsilon
_{A_{1}\cdots A_{2n+1}}\theta ^{A_{1}A_{2}}e^{A_{3}}\left(
R^{A_{4}A_{5}}+t^{2}(\theta
^{2})^{A_{4}A_{5}}+s^{2}e^{A_{4}}e^{A_{5}}\right) \times \cdots  \nonumber \\
&&\qquad \qquad \cdots \times \left( R^{A_{2n}A_{2n+1}}+t^{2}(\theta
^{2})^{A_{2n}A_{2n+1}}+s^{2}e^{A_{2n}}e^{A_{2n+1}}\right) \,,
\end{eqnarray}%
or in a tensorial notation,
\begin{eqnarray}
B_{2n} &=&-2n\,\sqrt{-h}\int\limits_{0}^{1}dt\,\int\limits_{0}^{t}ds\,\delta
_{\left[ j_{1}\cdots j_{2n-1}\right] }^{\left[ i_{1}\cdots i_{2n-1}\right]
}\,K_{i_{1}}^{j_{1}}\left( {\frac{1}{2}\,}%
R_{i_{2}i_{3}}^{j_{2}j_{3}}(h)-t^{2}K_{i_{2}}^{j_{2}}K_{i_{3}}^{j_{3}}+s^{2}%
\delta _{i_{2}}^{j_{2}}\delta _{i_{3}}^{j_{3}}\right) \times \cdots
\nonumber \\
&&\qquad \qquad \cdots \times \left( {\frac{1}{2}\,}%
R_{i_{2n-2}i_{2n-1}}^{j_{2n-2}j_{2n-1}}(h)-t^{2}K_{i_{2n-2}}^{j_{2n-2}}K_{i_{2n-1}}^{j_{2n-1}}+s^{2}\delta _{i_{2n-2}}^{j_{2n-2}}\delta _{i_{2n-1}}^{j_{2n-1}}\right) \,.
\label{B_2n}
\end{eqnarray}%
where the coupling constant takes the value $c_{2n}=\kappa $.

The explicit expression of the above boundary term can also be worked out
from an extension of a Chern-Simons density (called Transgression Form) for
the AdS group. This mathematical structure introduces an additional gauge
connection in the same homotopy class, such that the full action is truly
gauge-invariant \cite{Mora-Olea-Troncoso-Zanelli-TF}.

The on-shell variation of the complete action (\ref{I2n+1BC}) produces the
surface term
\begin{eqnarray}
\delta {\cal I}_{2n+1} &=&-2n\kappa \int\limits_{\partial
M_{2n+1}}\int\limits_{0}^{1}dt\ t\,\varepsilon _{a_{1}\cdots a_{2n}}\left(
\delta K^{a_{1}}e^{a_{2}}-K^{a_{1}}\delta e^{a_{2}}\right) \left(
R^{a_{3}a_{4}}-t^{2}K^{a_{3}}K^{a_{4}}+t^{2}e^{a_{3}}e^{a_{4}}\right) \nonumber \\
&&\qquad \qquad \cdots \times \left(
R^{a_{2n-1}a_{2n}}-t^{2}K^{a_{2n-1}}K^{a_{2n}}+t^{2}e^{a_{2n-1}}e^{a_{2n}}%
\right) \,,
\end{eqnarray}%
that, written in terms of tensors, becomes
\begin{eqnarray}
\delta {\cal I}_{2n+1} &=&2n\kappa \int\limits_{\partial
M_{2n+1}}d^{2n}x\, \sqrt{-h}\int\limits_{0}^{1}dt\ t\,\delta
_{\left[ j_{1}\cdots j_{2n}\right] }^{\left[ i_{1}\cdots
i_{2n}\right] } \left( \delta K_{i_{1}}^{j_{1}}\delta
_{i_{2}}^{j_{2}}+\frac{1}{2} \,K_{i_{1}}^{k}(h^{-1}\delta
h)_{k}^{j_{1}}\delta _{i_{2}}^{j_{2}}-\frac{1}{2
}\,K_{i_{1}}^{j_{1}}(h^{-1}\delta h)_{i_{2}}^{j_{2}}\right) \nonumber \\
&&\hspace{-1.1cm} \left( {\frac{1}{ 2}\,}
R_{i_{3}i_{4}}^{j_{3}j_{4}}(h)-t^{2}K_{i_{3}}^{j_{3}}K_{i_{4}}^{j_{4}}+t^{2}
\,\delta _{i_{3}}^{j_{3}}\delta _{i_{4}}^{j_{4}}\right)  \cdots
\left( {\frac{1}{2}\,}
R_{i_{2n-1}i_{2n}}^{j_{2n-1}j_{2n}}(h)-t^{2}K_{i_{2n-1}}^{j_{2n-1}}K_{i_{2n}}^{j_{2n}}+t^{2}\,\delta
_{i_{2n-1}}^{j_{2n-1}}\delta _{i_{2n}}^{j_{2n}}\right).
\label{varI_2n+1}
\end{eqnarray}
For an AAdS spacetime,
the metric expansion (\ref{gFG}) implies
\begin{equation}
K_{j}^{i}=\frac{1}{\ell }\,\delta _{j}^{i}-\frac{1}{\ell }\,\rho
\,(g_{(1)})_{j}^{i}+\cdots \,,  \label{K expansion}
\end{equation}%
where the indices are lowered and raised by $g_{(0)ij}$. So, we will
consider the condition
\begin{equation}
K_{j}^{i}=\frac{1}{\ell }\,\delta _{j}^{i}\,,  \label{K=delta}
\end{equation}%
such that
\begin{equation}
\delta K_{j}^{i}=0  \label{deltaK=0}
\end{equation}%
on the boundary, to cancel identically the different terms in the variation $%
\delta {\cal I}_{2n+1}$ \cite{Miskovic-Olea}.

It can be proved that the boundary term (\ref{B_2n}) renders the Euclidean
action finite and recovers the correct black hole thermodynamics for static
Chern-Simons-AdS solutions \cite{Banados-Teitelboim-Zanelli-continuedBH}. In
addition, the conserved quantities can be constructed as Noether charges
associated to asymptotic symmetries. However, it is clear from Eq.(\ref%
{varI_2n+1}) that this action does not lend itself to a clear definition of
a boundary stress tensor, as its variation contains pieces along $\delta
K_{j}^{i}$\ that are usually cancelled by a generalized Gibbons-Hawking
term. This might make difficult the holographic interpretation of this
method in the light of the AdS/CFT correspondence, where the boundary metric
is kept fixed at the boundary.

Because of the delicate point mentioned above, a note of caution is in order
here. The Dirichlet problem, defined as in Section \ref{Dirichlet problem},
does not really make sense for manifolds that are endowed with a conformal
boundary, as it is the case of AAdS spacetimes. Indeed, the leading order of
the expansion (\ref{gFG}) for the boundary metric $h_{ij}=g_{ij}/\rho $
makes a Dirichlet condition inappropriate for the variational problem
because of the divergence at $\rho =0$. Thus, one should fix the conformal
structure $g_{(0)ij}$ instead, and consider the addition of boundary terms
to cancel the divergences at the conformal boundary. It has been argued in
\cite{Papadimitriou-Skenderis} that these boundary terms are indeed the
Dirichlet counterterms, required originally by the regularization problem.
This reasoning reflects an interesting connection between the boundary terms
needed for a well-defined variation of the action and those that produce the
action regularization. It also resembles on the regularization scheme given
by Eq.(\ref{IregKT}), where the interplay between the variational principle
and the regularization problem is encoded in a single boundary term $B_{d}$.

The boundary condition (\ref{K=delta}) and its corresponding variation
simply correspond to the regular form of the Dirichlet condition on $%
g_{(0)ij}$. This is a consequence of the fact that, in AAdS spacetimes, the
leading order in Fefferman-Graham expansion for both the extrinsic curvature
$K_{ij}$ and the boundary metric $h_{ij}/\ell $ agree, what is no longer
true in the flat limit $\ell \rightarrow \infty $.\ By selecting regular
boundary conditions at $\rho =0$, one can be certain that no additional
divergences are introduced and, therefore, no extra counterterms are
required on top of the series (\ref{B_2n}). The compatibility of this
approach with keeping fixed $g_{(0)ij}$, together with the finiteness of the
variation of the action,\ strongly suggests that the holographic
reconstruction of the spacetime is already built-in in the Kounterterms
series.

In what follows, we combine both the intrinsic and the extrinsic
regularization mechanisms, in order to identify the Dirichlet counterterms
as the difference between the Kounterterms $B_{2n}$ and the generalized
Gibbons-Hawking term $\beta _{2n}$. First, we illustrate this procedure in
the five-dimensional case, where the action is
\begin{equation}
{\cal I}_{5}=I_{5}+\kappa \int\limits_{\partial M_{5}}d^{4}x\,B_{4}\,,
\end{equation}%
with
\begin{equation}
B_{4}=-\,\sqrt{-h}\delta _{\left[ j_{1}j_{2}j_{3}\right] }^{\left[
i_{1}i_{2}i_{3}\right] }\,K_{i_{1}}^{j_{1}}\left(
R_{i_{2}i_{3}}^{j_{2}j_{3}}(h)-K_{i_{2}}^{j_{2}}K_{i_{3}}^{j_{3}}+\frac{1}{3}%
\,\delta _{i_{2}}^{j_{2}}\delta _{i_{3}}^{j_{3}}\right) .
\end{equation}%
Now, let us simply insert the generalized Gibbons-Hawking term $\beta _{4}$
in a convenient manner,
\begin{equation}
{\cal I}_{5}=I_{5}+\kappa \int\limits_{\partial M_{5}}d^{4}x\,\beta
_{4}+\kappa \int\limits_{\partial M_{5}}d^{4}x\,\left( B_{4}-\beta
_{4}\right) \,,
\end{equation}%
such that the first two terms correspond to the Dirichlet action $%
I_{5}^{Dir} $ and will produce the finite stress tensor studied in Ref.\cite%
{Banados-Olea-Theisen}, plus two divergent terms
\begin{equation}
\delta I_{5}^{Dir}=\frac{1}{2}\int\limits_{\partial M_{5}}d^{4}x\,\sqrt{-g}%
\,T^{ij}\,\delta g_{ij}-\kappa \int\limits_{\partial M_{5}}d^{4}x\,\sqrt{-g}%
\,(g^{-1}\delta g)_{i}^{j}\left( \frac{8}{\rho ^{2}}\,\delta _{j}^{i}+\frac{1%
}{\rho }\,\delta _{\lbrack
jj_{1}j_{2}]}^{[ii_{1}i_{2}]}\,R_{i_{1}i_{2}}^{j_{1}j_{2}}(g)\right) \,.
\label{deltaI5Dh}
\end{equation}%
Then, we compute the difference $\left( B_{4}-\beta _{4}\right) $ as
\begin{equation}
\left( B_{4}-\beta _{4}\right) =\sqrt{-h}\,\delta _{\left[ j_{1}j_{2}j_{3}%
\right] }^{\left[ i_{1}i_{2}i_{3}\right] }\,K_{i_{1}}^{j_{1}}\left(
R_{i_{2}i_{3}}^{j_{2}j_{3}}(h)-\frac{1}{3}%
\,K_{i_{2}}^{j_{2}}K_{i_{3}}^{j_{3}}+\delta _{i_{2}}^{j_{2}}\delta
_{i_{3}}^{j_{3}}\right) \,,
\end{equation}%
and expanding the extrinsic curvature $K_{i}^{j}$ in the radial coordinate,
we realize that in the above relation, the divergent pieces do not depend on
$k_{i}^{j}$. The different contributions can be finally seen as the local
counterterms necessary to cancel the divergent terms in Eq.(\ref{deltaI5Dh}%
), that is,
\begin{equation}
{\cal L}_{4}=\kappa \left( B_{4}-\beta _{4}\right) =2\kappa \,\sqrt{-h}%
\left( 8+\delta _{\lbrack
j_{1}j_{2}]}^{[i_{1}i_{2}]}\,R_{i_{1}i_{2}}^{j_{1}j_{2}}(h)\right) +{\cal O}%
(1)\,.
\end{equation}%
The ${\cal O}(1)$ part left over at the boundary in the above difference,
\begin{eqnarray}
{\cal L}_{4}^{fin} &=&-\kappa \,\sqrt{-g}\,\delta _{\left[ j_{1}j_{2}j_{3}%
\right] }^{\left[ i_{1}i_{2}i_{3}\right] }\,k_{i_{1}}^{j_{1}}\left(
R_{i_{2}i_{3}}^{j_{2}j_{3}}(g)+k_{i_{2}}^{j_{2}}\delta
_{i_{3}}^{j_{3}}\right)  \nonumber \\
&=&\kappa \sqrt{-g}\,\left( \frac{1}{8}\,\delta _{\left[ j_{1}j_{2}j_{3}j_{4}%
\right] }^{\left[ i_{1}i_{2}i_{3}i_{4}\right] }%
\,R_{i_{1}i_{2}}^{j_{1}j_{2}}(g)\,R_{i_{3}i_{4}}^{j_{3}j_{4}}(g)+2\delta _{%
\left[ j_{1}j_{2}\right] }^{\left[ i_{1}i_{2}\right] }%
\,k_{i_{1}}^{j_{1}}k_{i_{2}}^{j_{2}}\right) \,,  \label{fin-4}
\end{eqnarray}%
corresponds to the Euler-Gauss-Bonnet invariant in four dimensions plus a
finite counterterm that does not contribute to the trace anomaly. (In the
last line, the equation of motion (\ref{CS-eom}), $E_{\rho }^{\rho }=0$, was
used.) This expression involves $k_{(0)ij}=g_{(1)ij}\,$, whose local piece
has a universal form in terms of the Ricci tensor $R_{(0)ij}$ for any
gravity theory with quadratic couplings in the curvature \cite%
{Imbimbo-Schwimmer-Theisen-Yankielowicz} (except for Chern-Simons \cite%
{Banados-Schwimmer-Theisen}). Then, in general, this term will give rise to
a quadratic combination of the curvature $R_{(0)kl}^{ij}$. This ambiguity is
even present in five-dimensional Einstein-Hilbert gravity, where one can
always add to the action quadratic terms in the curvature $R_{kl}^{ij}(h)$\
as scheme-dependent, finite counterterms that do not modify the Weyl anomaly
\cite{Balasubramanian-Kraus}.

The same trick can be done in higher odd dimensions, such that,
\begin{equation}
{\cal I}_{2n+1}=I_{2n+1}^{Dir}+\int\limits_{\partial M_{2n+1}}d^{2n}x\,{\cal %
L}_{2n}\,,
\end{equation}%
where
\begin{eqnarray}
{\cal L}_{2n} &=&\left( B_{2n}-\beta _{2n}\right) \\
&=&2n\kappa \,\sqrt{-h}\int\limits_{0}^{1}dt\,\int\limits_{t}^{1}ds\,\delta
_{\left[ j_{1}\cdots j_{2n-1}\right] }^{\left[ i_{1}\cdots i_{2n-1}\right]
}\,K_{i_{1}}^{j_{1}}\left( {\frac{1}{2}\,}%
R_{i_{2}i_{3}}^{j_{2}j_{3}}(h)-t^{2}K_{i_{2}}^{j_{2}}K_{i_{3}}^{j_{3}}+s^{2}%
\,\delta _{i_{2}}^{j_{2}}\delta _{i_{3}}^{j_{3}}\right) \times \cdots
\nonumber \\
&&\qquad \qquad \cdots \times \left( {\frac{1}{2}\,}%
R_{i_{2n-2}i_{2n-1}}^{j_{2n-2}j_{2n-1}}(h)-t^{2}K_{i_{2n-2}}^{j_{2n-2}}K_{i_{2n-1}}^{j_{2n-1}}+s^{2}\,\delta _{i_{2n-2}}^{j_{2n-2}}\delta _{i_{2n-1}}^{j_{2n-1}}\right) \,.
\label{explicitCT}
\end{eqnarray}%
In the expansion of the extrinsic curvature $K_{j}^{i}=\delta _{j}^{i}-\rho
k_{j}^{i}$ for the above expression, the divergent terms never contain $%
k_{j}^{i}$. Then, $k_{j}^{i}$ is only present in the finite piece and terms
that vanish as $\rho \rightarrow 0$. More explicitly, the expansion in $D=7$
and $D=9$ reads
\begin{eqnarray*}
{\cal L}_{6} &=&6\kappa \,\sqrt{-h}\left( 64+4\,\delta _{\lbrack
j_{1}j_{2}]}^{[i_{1}i_{2}]}\,R_{i_{1}i_{2}}^{j_{1}j_{2}}(h)+\frac{1}{4}%
\,\delta _{\lbrack
j_{1}j_{2}j_{3}j_{4}]}^{[i_{1}i_{2}i_{3}i_{4}]}%
\,R_{i_{1}i_{2}}^{j_{1}j_{2}}(h)\,R_{i_{3}i_{4}}^{j_{3}j_{4}}(h)\right) \,,
\\
{\cal L}_{8} &=&24\kappa \,\sqrt{-h}\left( 768+32\,\delta _{\lbrack
j_{1}j_{2}]}^{[i_{1}i_{2}]}\,R_{i_{1}i_{2}}^{j_{1}j_{2}}(h)+\delta _{\lbrack
j_{1}j_{2}j_{3}j_{4}]}^{[i_{1}i_{2}i_{3}i_{4}]}%
\,R_{i_{1}i_{2}}^{j_{1}j_{2}}(h)R_{i_{3}i_{4}}^{j_{3}j_{4}}(h)+\right. \\
&&\left. +\frac{1}{24}\,\delta _{\lbrack
j_{1}j_{2}j_{3}j_{4}j_{5}j_{6}]}^{[i_{1}i_{2}i_{3}i_{4}i_{5}i_{6}]}%
\,R_{i_{1}i_{2}}^{j_{1}j_{2}}(h)R_{i_{3}i_{4}}^{j_{3}j_{4}}(h)R_{i_{5}i_{6}}^{j_{5}j_{6}}(h)\,\right) ,
\end{eqnarray*}%
up to a finite term of the type (\ref{fin-4}). The above examples show the
agreement with the counterterms obtained from the direct integration of
Dirichlet variation, Eq.(\ref{ct-CS}). Due to the lack of dependence on $%
k_{j}^{i}$, we might take directly $k_{j}^{i}=0$ into the general expression
for the counterterms (\ref{explicitCT}), to find explicitly the terms in the
Lovelock-type series
\begin{equation}
{\cal L}_{2n}=2n\kappa \,\sqrt{-h}\,\sum\limits_{p=0}^{n-1}\left(
\begin{array}{c}
n-1 \\
p%
\end{array}%
\right) \frac{d_{p}}{2^{p}}\,\delta _{\left[ j_{1}...j_{2p}\right] }^{\left[
i_{1}...i_{2p}\right] }%
\,R_{i_{1}i_{2}}^{j_{1}j_{2}}(h)...R_{i_{2p-1}i_{2p}}^{j_{2p-1}j_{2p}}(h)\,,
\label{Lct}
\end{equation}%
where the coefficients are evaluated as
\begin{eqnarray}
d_{p} &=&\left( 2n-2p\right)
!\int\limits_{0}^{1}dt\int\limits_{t}^{1}ds\left( s^{2}-t^{2}\right) ^{n-1-p}
\nonumber \\
&=&4^{n-p-1}(n-p-1)!^{2}\,.
\end{eqnarray}%
In summary, the difference between the Kounterterms $B_{2n}$ and the
generalized Gibbons-Hawking term $\beta _{2n}\,$ depends on $K_{j}^{i}$ and
might be even non-local. But, surprisingly, this procedure generates the
series of local Dirichlet counterterms (\ref{ct-CS}).

%%%%%%%%%%%%%%%%%%%%%%%%%%%%%%%%%%%%%%%%%%%%%%%%%%%%%%%%%%%%%%%%%%%%%%%%%%%%%%%%%%%

\subsection{Born-Infeld-AdS}

A mechanism to regularize the conserved quantities in Born-Infeld-AdS
gravity in $D=2n$ was discussed in Ref.\cite%
{Aros-Contreras-Olea-Troncoso-Zanelli-2n}, where it was proposed to add the $%
2n$-dimensional Euler term
\begin{eqnarray}
{\cal E}_{2n} &=&\varepsilon _{A_{1}\cdots A_{2n}}\,\hat{R}%
^{A_{1}A_{2}}\cdots \hat{R}^{A_{2n-1}A_{2n}}  \nonumber \\
&=&-\frac{1}{2^{n}}\,d^{2n}x\,\sqrt{-G}\,\delta _{\left[ \mu _{1}\cdots \mu
_{2n}\right] }^{\left[ \mu _{1}\cdots \mu _{2n}\right] }\,\hat{R}_{\mu
_{1}\mu _{2}}^{\mu _{1}\mu _{2}}\cdots \hat{R}_{\mu _{2n-1}\mu _{n}}^{\mu
_{2n-1}\mu _{2n}}
\end{eqnarray}%
to the bulk action (\ref{BI-bulk}). This is a topological invariant that
does not modify the field equations but gives a non-trivial contribution to
the Noether current. The coupling constant in front of ${\cal E}_{2n}$ is
adjusted proceeding in the following way: let us consider the action $%
I_{2n}+\alpha \int_{M_{2n}}{\cal E}_{2n}$ ($\alpha $ is an
arbitrary coupling constant) whose on-shell variation produces the
surface term
\begin{eqnarray}
\delta \left( I_{2n}+\alpha \int\limits_{M_{2n}}{\cal
E}_{2n}\right) &=&n\int\limits_{\partial M_{2n}}\varepsilon
_{A_{1}\cdots A_{2n}}\,\delta \omega ^{A_{1}A_{2}}\times  \nonumber \\
&& \times\left[ \kappa \left( \hat{R}^{A_{3}A_{4}}+\frac{1}{\ell
^{2}}\,e^{A_{3}}e^{A_{4}}\right) \cdots \left(
\hat{R}^{A_{2n-1}A_{2n}}+
\frac{1}{\ell ^{2}}\,e^{A_{2n-1}}e^{A_{2n}}\right) +\right.   \nonumber \\
&&\qquad \qquad +\left. \left( \alpha -\kappa \right) \,\hat{R}
^{A_{3}A_{4}}\cdots \hat{R}^{A_{2n-1}A_{2n}}\right] \,.
\end{eqnarray}
Therefore, demanding the spacetime to be asymptotically locally
AdS, i.e.,
\begin{equation}
\hat{R}_{\mu \nu }^{\alpha \beta }=-\frac{1}{\ell ^{2}}\,\delta _{\lbrack
\mu \nu ]}^{[\alpha \beta ]}  \label{ALAdS}
\end{equation}%
at the boundary, the action is stationary on-shell only if $\alpha =\kappa $%
. This comes as a natural generalization of a strategy used for
Einstein-Hilbert-AdS in any even dimension \cite%
{Aros-Contreras-Olea-Troncoso-Zanelli-2n,Aros-Contreras-Olea-Troncoso-Zanelli-4}%
.

In this way, the total action is
\begin{equation}
{\cal I}_{2n}=I_{2n}+\kappa \int\limits_{M_{2n}}{\cal E}_{2n}\,,
\end{equation}%
that takes the more compact form ($\ell =1$)
\begin{eqnarray}
{\cal I}_{2n} &=&\kappa \int\limits_{M_{2n}}\varepsilon _{A_{1}\cdots
A_{2n}}\,\left( \hat{R}^{A_{1}A_{2}}+e^{A_{1}}e^{A_{n}}\right) \cdots \left(
\hat{R}^{A_{2n-1}A_{2n}}+e^{A_{2n-1}}e^{A_{2n}}\right)  \nonumber \\
&=& -\kappa \int\limits_{M_{2n}}d^{2n}x\,\sqrt{-h}\,\delta
_{\left[ \nu _{1}\cdots \nu _{2n}\right] }^{\left[ \mu _{1}
\cdots\mu _{2n}\right] }\,\left( \frac{1}{2}\,\hat{R}_{\mu _{1}\mu
_{2}}^{\nu _{1}\nu _{2}}+\delta _{\mu _{1}}^{\nu _{1}}\delta _{\mu
_{2}}^{\nu _{2}}\right)\times\cdots\nonumber \\
&&\qquad\qquad\cdots \times \left( \frac{1}{2}\,\hat{R}_{\mu
_{2n-1}\mu _{2n}}^{\nu _{2n-1}\nu _{2n}}+\delta _{\mu
_{2n-1}}^{\nu _{2n-1}}\delta _{\mu _{2n}}^{\nu _{2n}}\right) \,.
\label{BI+E}
\end{eqnarray}%
For the purpose of comparison with the Dirichlet counterterms, it is
convenient to use the Euler theorem
\begin{equation}
\int\limits_{M_{2n}}d^{2n}x\,{\cal E}_{2n}=\left( -4\pi \right)
^{n}\,n!\,\chi (M_{2n})+\int\limits_{\partial M_{2n}}d^{2n-1}xB_{2n-1}\,,
\label{Euler term}
\end{equation}%
to obtain the equivalence to a Kounterterms series, that is, a boundary term
that depends on the extrinsic curvature $K_{j}^{i}$ and that is given by
\cite{OleaJHEP}
\begin{eqnarray}
B_{2n-1} &=&2n\,\sqrt{-h}\int\limits_{0}^{1}dt\,\delta _{\lbrack i_{1}\cdots
i_{2n-1}]}^{[j_{1}\cdots j_{2n-1}]}\,K_{j_{1}}^{i_{1}}\left( \frac{1}{2}%
\,R_{j_{2}j_{3}}^{i_{2}i_{3}}(h)-t^{2}\,K_{j_{2}}^{i_{2}}K_{j_{3}}^{i_{3}}%
\right) \times \cdots  \nonumber \\
&&\qquad \qquad \cdots \times \left( \frac{1}{2}%
\,R_{j_{2n-2}j_{2n-1}}^{i_{2n-2}i_{2n-1}}(h)-t^{2}%
\,K_{j_{2n-2}}^{i_{2n-2}}K_{j_{2n-1}}^{i_{2n-1}}\right) ,
\label{Kounter_2n-1}
\end{eqnarray}%
with a coupling constant $c_{2n-1}=\kappa $.

Performing a similar procedure as in the Chern-Simons case, we add and
subtract the generalized Gibbons-Hawking term into the action
\begin{equation}
{\cal I}_{2n}=I_{2n}+\kappa \int\limits_{\partial
M_{2n}}d^{2n-1}x\,B_{2n-1}\,,  \label{cal I}
\end{equation}%
in order to identify the divergent parts,
\begin{equation}
{\cal I}_{2n}=I_{2n}^{Dir}+\kappa \int\limits_{\partial
M_{2n}}d^{2n-1}x\,\left( B_{2n-1}-\beta _{2n-1}\right) \,.
\end{equation}%
The first term in the above expression corresponds to the Dirichlet action,
and the second part can be cast into the parametric integration
\begin{eqnarray}
\left( B_{2n-1}-\beta _{2n-1}\right) &=&2n\,\sqrt{-h}\int\limits_{0}^{1}dt\,%
\delta _{\left[ j_{1}\cdots j_{2n-1}\right] }^{\left[ i_{1}\cdots i_{2n-1}%
\right] }\,K_{i_{1}}^{j_{1}}\left( {\frac{1}{2}\,}%
R_{i_{2}i_{3}}^{j_{2}j_{3}}(h)-t^{2}K_{i_{2}}^{j_{2}}K_{i_{3}}^{j_{3}}+%
\delta _{i_{2}}^{j_{2}}\delta _{i_{3}}^{j_{3}}\right) \times \cdots
\nonumber \\
&&\cdots \times \left( {\frac{1}{2}\,}%
R_{i_{2n-2}i_{2n-1}}^{j_{2n-2}j_{2n-1}}(h)-t^{2}K_{i_{2n-2}}^{j_{2n-2}}K_{i_{2n-1}}^{j_{2n-1}}+\delta _{i_{2n-2}}^{j_{2n-2}}\delta _{i_{2n-1}}^{j_{2n-1}}\right) \,,
\end{eqnarray}%
using the identity (\ref{binomial}).

Expanding the above formula using the relations (\ref{rescaled h}-\ref%
{rescaled var}) and the determinant of the boundary metric
\begin{equation}
\sqrt{-h}=\frac{\sqrt{-g}}{\rho ^{n-\frac{1}{2}}}\,,
\end{equation}%
we notice that the divergent terms do not depend on $k_{j}^{i}$. As a
consequence, they can be computed by setting $k_{j}^{i}=0$ and performing
the integration in the parameter $t$, so that we have
\begin{equation}
\kappa \left( B_{2n-1}-\beta _{2n-1}\right) =2n\kappa \,\sqrt{-h}%
\sum\limits_{p=0}^{n-1}\left(
\begin{array}{c}
n-1 \\
p%
\end{array}%
\right) \frac{d_{p}}{2^{p}}\,\delta _{\left[ j_{1}...j_{2p}\right] }^{\left[
i_{1}...i_{2p}\right] }\,R_{i_{1}i_{2}}^{j_{1}j_{2}}(h)\cdots
R_{i_{2p-1}i_{2p}}^{j_{2p-1}j_{2p}}(h)\,,
\end{equation}%
where the coefficients are
\begin{eqnarray*}
d_{p} &=&\left( 2n-2p-1\right) !\,\int\limits_{0}^{1}dt\,\left(
1-t^{2}\right) ^{n-p-1} \\
&=&4^{n-p-1}(n-p-1)!^{2}\,.
\end{eqnarray*}%
They can be identified, up to ${\cal O}(\rho ^{-3/2})$, with the Dirichlet
counterterms (\ref{CT-BI}),
\begin{equation}
\kappa \left( B_{2n-1}-\beta _{2n-1}\right) ={\cal L}_{2n-1}+\frac{n\kappa }{%
2^{n-2}}\,\frac{\sqrt{-g}}{\rho ^{\frac{1}{2}}}\,\delta _{\left[
j_{1}...j_{2n-2}\right] }^{\left[ i_{1}...i_{2n-2}\right] }%
\,R_{i_{1}i_{2}}^{j_{1}j_{2}}(g)\cdots
R_{i_{2n-31}i_{2n-2}}^{j_{2n-3}j_{2n-2}}(g)\,.  \label{B-Beta}
\end{equation}%
In both Chern-Simons and Born-Infeld AdS gravities, if one considers
flat-boundary spacetimes ($R_{kl}^{ij}(h)=0$), the Dirichlet counterterms
series (Eqs.(\ref{ct-CS}) and (\ref{CT-BI}), respectively) reduces to a
single counterterm proportional to the induced volume of the boundary.
Though this corresponds to a very particular case, this term is yet enough
to regularize the conserved charges for horizonless extended solutions in
these theories \cite{Dehghani-Bostani-Sheikhi}.

The last term of the Eq.(\ref{B-Beta}) contributes to the {\em finite} part
of the stress tensor and, as it can be seen from the variation of the action
(\ref{BI+E}) as\footnote{%
We have neglected a term along $\delta \omega ^{ab}$, that can be expressed
in terms of the variation of Christoffel symbol $\Gamma _{jk}^{i}(h)=\Gamma
_{jk}^{i}(g)$, because it is of order ${\cal O}(\sqrt{\rho })$.}
\begin{eqnarray}
\delta {\cal I}_{2n} &=&2n\kappa \int\limits_{\partial M_{2n}}d^{2n-1}x\,%
\sqrt{-h}\,\delta _{\left[ j_{1}\cdots j_{2n-1}\right] }^{\left[ i_{1}\cdots
i_{2n-1}\right] }\left( \delta K_{i_{1}}^{j_{1}}+\frac{1}{2}%
\,K_{i_{1}}^{k}(h^{-1}\delta h)_{k}^{j_{1}}\right)\times\nonumber \\
&&\times \left( {\frac{1}{2}\,}
R_{i_{2}i_{3}}^{j_{2}j_{3}}(h)-K_{i_{2}}^{j_{2}}K_{i_{3}}^{j_{3}}+\delta
_{i_{2}}^{j_{2}}\delta _{i_{3}}^{j_{3}}\right) \times \cdots  \nonumber \\
&&\qquad\cdots \times \left( {\frac{1}{2}\,}
R_{i_{2n-2}i_{2n-1}}^{j_{2n-2}j_{2n-1}}(h)-K_{i_{2n-2}}^{j_{2n-2}}K_{i_{2n-1}}^{j_{2n-1}}+\delta
_{i_{2n-2}}^{j_{2n-2}}\delta _{i_{2n-1}}^{j_{2n-1}}\right) .
\label{aux_var}
\end{eqnarray}%
Indeed, counting powers of $\rho $, the term in Eq.(\ref{aux_var}) along $%
\delta K_{i}^{j}=-\rho \,\delta k_{i}^{j}$ vanishes in the limit $\rho
\rightarrow 0$, such that the stress tensor has the form
\begin{eqnarray}
T_{j}^{i}(h) &=&2n\kappa \,\delta _{\left[ jj_{2}\cdots j_{2n-1}\right] }^{%
\left[ ki_{2}\cdots i_{2n-1}\right] }K_{k}^{i}\left( {\frac{1}{2}\,}%
R_{i_{2}i_{3}}^{j_{2}j_{3}}(h)-K_{i_{2}}^{j_{2}}K_{i_{3}}^{j_{3}}+\delta
_{i_{2}}^{j_{2}}\delta _{i_{3}}^{j_{3}}\right) \times \cdots  \nonumber \\
&&\cdots \times \left( {\frac{1}{2}\,}%
R_{i_{2n-2}i_{2n-1}}^{j_{2n-2}j_{2n-1}}(h)-K_{i_{2n-2}}^{j_{2n-2}}K_{i_{2n-1}}^{j_{2n-1}}+\delta _{i_{2n-2}}^{j_{2n-2}}\delta _{i_{2n-1}}^{j_{2n-1}}\right) .
\label{T(h)}
\end{eqnarray}

The corresponding conserved quantities are constructed assuming that the
boundary submanifold can be foliated in time-like ADM form
\begin{equation}
h_{ij}\,dx^{i}dx^{j}=-N_{\Sigma }^{2}(t)\,dt^{2}+\sigma _{nm}\left( d\varphi
^{n}+N_{\Sigma }^{n}dt\right) \left( d\varphi ^{m}+N_{\Sigma }^{m}dt\right)
\,,
\end{equation}%
with the coordinates $x^{i}=\left( t,\varphi ^{m}\right) $ and defined by
the time-like unit normal $n_{i}=(-N_{\Sigma },\vec{0})$. The charges are
then given as the integration on $\Sigma $ (the boundary of spatial section)
that is parametrized by $\varphi ^{m}$,
\begin{equation}
Q(\xi )=\int\limits_{\Sigma }d^{2n-2}\varphi \,\sqrt{\sigma }%
\,T_{j}^{i}(h)\,\xi ^{j}n_{i}\,,  \label{Q}
\end{equation}%
where $\sigma $ denotes the determinant of the metric $\sigma _{nm}$ (that
satisfies $\sqrt{-h}=N_{\Sigma }\sqrt{\sigma }$) and $\xi ^{i}$ is an
asymptotic Killing vector. It can be verified, with the help of some of the
identities extensively used above, that the conserved quantity (\ref{Q})
agrees with the charge obtained by the Noether theorem in differential forms
language \cite{Aros-Contreras-Olea-Troncoso-Zanelli-2n}, and provides the
correct mass for Born-Infeld-AdS black holes \cite%
{Banados-Teitelboim-Zanelli-continuedBH,Cai-Soh}.

Expanding the form of Eq.(\ref{T(h)}), we notice that a finite stress tensor
can be obtained multiplying $T_{j}^{i}(h)$ by a suitable factor
\begin{equation}
T_{j}^{i}=\lim_{\rho \rightarrow 0}\frac{1}{\rho ^{\frac{d-1}{2}}}%
\,T_{j}^{i}(h)\,,
\end{equation}%
and can be written as
\begin{eqnarray}
T_{j}^{i} &=&n\kappa \,\delta _{\left[ jj_{1}\cdots j_{2n-2}\right] }^{\left[
ii_{1}\cdots i_{2n-2}\right] }\,(g^{-1}\delta g)_{i}^{j}\,\left( {\ \frac{1}{%
2}\,}R_{i_{1}i_{2}}^{j_{1}j_{2}}+2\,k_{i_{1}}^{j_{1}}\delta
_{i_{2}}^{j_{2}}\right) \cdots \left( {\frac{1}{2}\,}%
R_{i_{2n-3}i_{2n-2}}^{j_{2n-3}j_{2n-2}}+2\,k_{i_{2n-3}}^{j_{2n-3}}\delta
_{i_{2n-2}}^{j_{2n-2}}\right)   \nonumber \\
&=&(T_{j}^{i})_{Dir}+\frac{n\kappa }{2^{n-2}}\,\delta _{\left[ jj_{1}\cdots
j_{2n-2}\right] }^{\left[ ii_{1}\cdots i_{2n-2}\right] }%
\,R_{i_{1}i_{2}}^{j_{1}j_{2}}\cdots
R_{i_{2n-3}i_{2n-2}}^{j_{2n-3}j_{2n-2}}\,,  \label{TgBI}
\end{eqnarray}%
where only the first term in $K_{k}^{i}=\delta _{k}^{i}-\rho k_{k}^{i}$\ of
the first line of Eq.(\ref{T(h)}) contributes to the stress tensor. Using
the components $E_{\rho }^{\rho }$ of the equations of motion (\ref{EOMBI}),
one can prove that the trace of the above stress tensor, as expected,
vanishes identically.

The first piece of the expression (\ref{TgBI}), $(T_{j}^{i})_{Dir}$, can be
read off from the variation of the Dirichlet action (\ref{TDirBI}). This
argument shows the consistency between Dirichlet counterterms and
Kounterterms also at the level of the regularized stress tensors, as they
differ at most by a finite term.

%%%%%%%%%%%%%%%%%%%%%%%%%%%%%%%%%%%%%%%%%%%%%%%%%%%%%%%%%%%%%%%%%%%%%%%%%%%%%%%%%%%
%%%%%%%%%%%%%%%%%%%%%%%%%%%%%%%%%%%%%%%%%%%%%%%%%%%%%%%%%%%%%%%%%%%%%%%%%%%%%%%%%%%

\section{Conclusions}

In this paper, we have performed the first direct comparison between
Dirichlet regularization of AdS gravity and Kounterterms prescription in two
particular Lovelock theories that feature a symmetry enhancement. The
remarkable agreement of the counterterms that produce the divergences
cancellation in the action and stress tensor, indicates that a similar
property should appear also in other Lovelock gravities with AdS asymptotics.

At this level, we simply conjecture that the Dirichlet counterterms in any
Lovelock-AdS theory should be generated as the difference\footnote{%
Once again, ${\cal O}(1)\,$\ represents a finite term that, when $d$\ is
even, does not change the trace anomaly. In turn, just because of an
argument of dimensionality, when $d$\ is odd the extra term will be
proportional to $1/\sqrt{\rho }$\ that corresponds to a finite extra
contribution to the stress tensor.}
\begin{equation}
c_{d}B_{d}-\kappa \beta _{d}={\cal L}_{d}+{\cal O}(1)\,,  \label{DirCT}
\end{equation}%
though a final proof of it might be more involved than in the cases treated
here.

%%%%%%%%%%%%%%%%%%%%%%%%%%%%%%%%%%%%%%%%%%%%%%%%%%%%%%%%%%%%%%%%%%%%%%%%%%%%%%%%%%%
%%%%%%%%%%%%%%%%%%%%%%%%%%%%%%%%%%%%%%%%%%%%%%%%%%%%%%%%%%%%%%%%%%%%%%%%%%%%%%%%%%%

\section*{Acknowledgments}

We would like to thank M. Ba\~{n}ados and S. Theisen for enlightening
discussions at an early stage of this work. We are also grateful to S.
Detournay, D. Klemm and G. Kofinas for helpful conversations. O.M. is
supported by the PUCV through the program Investigador Joven 2007. O.M.
would like to thank the INFN Sezione di Milano for hospitality during the
preparation of this work. The work of R.O. is supported by INFN. R.O. also
thanks the organizers of the Workshop {\em String and M theory approaches to
particle physics and cosmology} at Galileo Galilei Institute, Florence, J.
Edelstein for hospitality at USC, Santiago de Compostela and G. Barnich for
hospitality at ULB, Brussels.

%%%%%%%%%%%%%%%%%%%%%%%%%%%%%%%%%%%%%%%%%%%%%%%%%%%%%%%%%%%%%%%%%%%%%%%%%%%%%%%%%%%
%%%%%%%%%%%%%%%%%%%%%%%%%%%%%%%%%%%%%%%%%%%%%%%%%%%%%%%%%%%%%%%%%%%%%%%%%%%%%%%%%%%

\appendix

\section{Useful identities}

The totally-antisymmetric Kronecker delta of rank $m$ is defined as the
determinant
\begin{equation}
\delta _{\left[ \mu _{1}\cdots \mu _{m}\right] }^{\left[ \nu _{1}\cdots \nu
_{m}\right] }:=\left\vert
\begin{array}{cccc}
\delta _{\mu _{1}}^{\nu _{1}} & \delta _{\mu _{1}}^{\nu _{2}} & \cdots &
\delta _{\mu _{1}}^{\nu _{m}} \\
\delta _{\mu _{2}}^{\nu _{1}} & \delta _{\mu _{2}}^{\nu _{2}} &  & \delta
_{\mu _{2}}^{\nu _{m}} \\
\vdots &  & \ddots &  \\
\delta _{\mu _{m}}^{\nu _{1}} & \delta _{\mu _{m}}^{\nu _{2}} & \cdots &
\delta _{\mu _{m}}^{\nu _{m}}%
\end{array}%
\right\vert \,.
\end{equation}%
A contraction of $k$ indices in the above Kronecker delta produces a delta
of order $m-k$,
\begin{equation}
\delta _{\left[ \mu _{1}\cdots \mu _{k}\cdots \mu _{m}\right] }^{\left[ \nu
_{1}\cdots \nu _{k}\cdots \nu _{m}\right] }\,\delta _{\nu _{1}}^{\mu
_{1}}\cdots \delta _{\nu _{k}}^{\mu _{k}}=\frac{\left( N-m+k\right) !}{%
\left( N-m\right) !}\;\delta _{\left[ \mu _{k+1}\cdots \mu _{m}\right] }^{%
\left[ \nu _{k+1}\cdots \nu _{m}\right] }\,,\qquad (1\leq k\leq m\leq N)\,,
\end{equation}%
where $N$ is the range of indices.

A useful identity that has been employed in the paper involves the binomial
expansion given in an integral form,
\begin{equation}
\left( a+b\right) ^{p}=a^{p}+p\,b\int\limits_{0}^{1}du\,\left( a+ub\right)
^{p-1}\,,\qquad p\geq 1\,.  \label{binomial}
\end{equation}%
Other two integral representations of a binomial often used in the text are
\begin{eqnarray}
\int\limits_{0}^{1}dt\,\left[ a+(2p+1)\,t^{2}b\right] \left(
a+t^{2}b^{2}\right) ^{p-1} &=&\left( a+b\right) ^{p}\,,\qquad p\geq 1\,, \\
\int\limits_{0}^{1}dt\,2t\left[ a+(p+1)\,t^{2}b\right] \left(
a+t^{2}b\right) ^{p-1} &=&(a+b)^{p}\,,\qquad p\geq 1\,.
\end{eqnarray}

%%%%%%%%%%%%%%%%%%%%%%%%%%%%%%%%%%%%%%%%%%%%%%%%%%%%%%%%%%%%%%%%%%%%%%%%%%%%%%%%%%%

\end{document}